\documentclass{appolb}
\usepackage{epsfig,myext}

\begin{document}
\title{Prospect for the Higgs searches with the ATLAS detector%
\thanks{Presented at Cracow Epiphany Conference on Hadron Interactions at the Dawn of the LHC,
dedicated to memory of J. Kwiecinski, Krak\' ow, Poland, 5-7 January 2009.} \\
}
\author{Elzbieta RICHTER-WAS 
\thanks{Supported in part by the RTN European Programme MRTN-CT-2006-035505 (HEPTOOLS, Tools and Precision 
Calculations for Physics Discoveries at Colliders) and by Polish Ministry of Science and Higher Education
 - 153/6PR UE/2007/7.} \\
       on behalf of the ATLAS Collaboration
\address{Institute of Physics, Jagellonian University,\\ Reymonta 4, 30-059~Krak\' ow, Poland\\
         Institute of Nuclear Physics Polish Academy of Sciences,\\ Radzikowskiego 152, 31-342~Krak\' ow, Poland}
}
\maketitle
\begin{abstract}
The investigation of the electroweak symmetry breaking is one of the primary tasks of the 
experiments at the CERN Large Hadron Collider (LHC). The potential of the ATLAS experiment 
for the discovery of the Higgs boson(s) in Standard Model and Minimal Supersymmetric 
Standard Model is presented, with emphasis on studies which have been completed recently. 
\end{abstract}
  
\section{Introduction}
The Large Hadron Collider at CERN will play an important role in the investigation of fundamental questions
of particle physics. While the Standard Model of electroweak \cite{EWModel} and strong \cite{QCDModel}
interactions is in excellent agreement with the numerous experimental measurements, 
the dynamics responsible for the electroweak symmetry breaking 
are still unknown. 
Within the Standard Model, the Higgs mechanism \cite{SMHiggs} is invoked to break the electroweak symmetry.
A doublet of complex of scalar fields is introduced, of which a single neutral scalar particle, the Higgs boson,
remains after symmetry breaking. Many extensions of this minimal version of the Higgs sector have been proposed.
Mostly discussed scenario is the one with two complex Higgs doublets, as realised in the Minimal Supersymmetric 
Standard Model (MSSM) \cite{SM&MSSM}, resulting in five observable Higgs bosons, three  electrically 
neutral (h,H,A) and two charged (H$^{\pm}$). At tree level their properties like masses, widths and 
branching fractions can be predicted in terms of only two parameters, typically chosen to be the mass of the 
CP-odd Higgs boson, $m_A$, and the tangent of the ratio of the vacuum expectation values of the two Higgs 
doublets, $\tan \beta$. 

Within the Standard Model, the Higgs boson is the only particle which has not been discovered so far. 
On the basis of theoretical knowledge, the Higgs sector in the Standard Model remains largely unconstrained.
While there is no direct prediction for the mass of the Higgs boson, an upper limit of  $\sim$ 1 TeV can 
be inferred from unitarity arguments \cite{UnitStabil}. Further constraints can be derived under the assumption that the 
Standard Model is valid only up to a cutoff energy scale $\Lambda$, beyond which new physics becomes relevant.
The requirement that the electroweak vacuum is stable and that the Standard Model remains perturbative allows to set 
upper and lower bounds on the Higgs boson mass \cite{UnitStabil}. For the cutoff scale  $\Lambda$ of the order 
of the Planck mass ($\tilde 1.22 \times 10^{19}$ GeV), the Higgs boson mass is required to be in the range 130 $< m_{H} <$ 180 GeV. 
If the new physics  appears at the lower mass scales, the bound becomes weaker, e.g. for  $\Lambda$ = 1 TeV the 
Higgs boson mass is constrained to be 50  $< m_{H} < $ 800 GeV.

The direct search at the $e^{+}e^{-}$ collider LEP has led to lower bound on its mass of 114.4 GeV at 95\% C.L.
\cite{LEPlimit}.
Indirectly, high precision electroweak data constrain the mass of the Higgs boson via their sensitivity to
loop corrections. Assuming the overall validity of the Standard Model, a global fit \cite{EWFit} to all electroweak 
data leads to the 95\% C.L. limit $m_{H} < 154$ GeV. The 95\% C.L. lower limit obtained from LEP is not used in the 
determination of this limit. Including it, the limit increases to 185 GeV. With the recently announced preliminary results 
from direct searches at Tevatron, with 2.0-3.6 $fb^{-1}$ of data analysed at CDF, and 0.9-4.2  $fb^{-1}$ at D0,  
the 95\% C.L. upper limits on the Higgs boson production are a factor of 2.5 (0.86) times the SM 
cross-section for a Higgs boson mass of $m_{H}=$ 115 (165) GeV. The mass range excluded at
95\% C.L. of a Standard Model Higgs boson has been extended to 160 $< m_{H} < 170$ GeV \cite{Tevatron}. 

The direct search at the $e^{+}e^{-}$ collider LEP has led also to limits in case of the MSSM Higgs 
bosons \cite{LEPlimit}. Values of $\tan \beta$ close to one and low $m_A$ ranges are excluded, see Ref.~\cite{MSSMLEP}. 
The Tevatron experiments
are reaching sensitivity \cite{MSSMChrgatTeV} to search for the charged Higgs boson in decay of $t \bar t$ 
pair with subsequent decay $H^{\pm} \to \tau \nu$ or direct production of H$^{\pm}$ and decay to $tb$. 
The reach of searches at the Tevatron \cite{MSSMatTeV} for neutral Higgs decaying into $\tau$-pair pair, 
interpreted in the MSSM context, extends the 95\% exclusion region below $\tan \beta$ = 40 for  
$m_A = 120 \div 160$ GeV.

\newpage 
In the following, I will discuss the potential for Higgs boson searches at the Large Hadron Collider with
the ATLAS experiment, focusing on  prospects for the initial
period, i.e. 10 $\div$ 30 fb$^{-1}$ of integrated luminosity at 14 TeV centre-of-mass collision,
and on analyses which have been recently revisited and published in
Ref.~\cite{CERN-OPEN-2008-020}. 

This new publication contains significant improvements in experimental methods and better understood
features of reconstruction algorithms than the previous work~\cite{ATLASTDR}. 
They are partially based on test beam measurements of various detector components or 
rely on extensive Monte Carlo simulations of the detailed detector response with as built detector
geometry \cite{ATLASDetPaper}. Here I will show final results of the analyses while I would like to 
address the reader to Ref.~\cite{ATLASDetPaper} and Ref.~\cite{CERN-OPEN-2008-020}  for the discussion 
of the expected detector performance for triggering 
given topologies or reconstructing physics objects (electrons, photons, muons, taus).

\section{Standard Model Higgs bosons at hadron colliders}

At hadron colliders the Higgs bosons can be produced via four different processes:
\begin{itemize}
\item gluon fusion, $gg \to H$, which is mediated at the lowest order by a heavy top-quark loop;
\item vector boson fusion (VBF), $qq \to qq H$;
\item associated production of the Higgs boson with weak gauge bosons, $qq \to W/Z H$;
\item associated Higgs boson production with heavy quarks, $gg, q\bar q \to t \bar t H$, 
$gg, q\bar q \to b \bar b H$ (and $gb \to bH$)
\end{itemize}

For all production processes higher-order QCD corrections have been calculated; in particular significant 
progress has been made over the last few years in calculation of QCD corrections for gluon-fusion and
for the associated $t \bar t H$ and $b \bar b H$ production processes, 
for a recent review see \cite{Jakobs-EurJPhysC}.
The production cross-section for the different processes is shown in Fig.~\ref{FS2.1} (left), 
following recent estimates \cite{CERN-OPEN-2008-020} with NLO accuracy whenever available. 
Due to impressive progress in the calculation of the higher-order QCD corrections for signal and backgrounds
over the past few years, the LHC physics studies have started to use these calculations.

The branching fractions of the Standard Model Higgs boson are shown in Fig.~\ref{FS2.1} (right), 
as a function of the Higgs boson mass. When kinematically accessible, decays of the Standard Model Higgs boson
into vector boson pairs $WW$ or $ZZ$ dominate over other decay modes. Above the kinematical threshold,
the branching fraction to $t \bar t$ can reach 20\%. All other fermionic decay modes are only relevant 
for the Higgs boson masses below $2 m_W$, with $H \to b \bar b$ dominating below 140 GeV. 
The branching fraction for $H \to \tau \tau$ reaches up to about 8\% at the Higgs boson masses between 100-120 GeV. 
Decays into photon pairs, which are of interest due to their relatively clean signature, can reach branching 
fraction of up to $2 \times 10^{-3}$ at low Higgs boson masses.

\begin{Fighere}
\begin{center}
{
  \epsfig{file=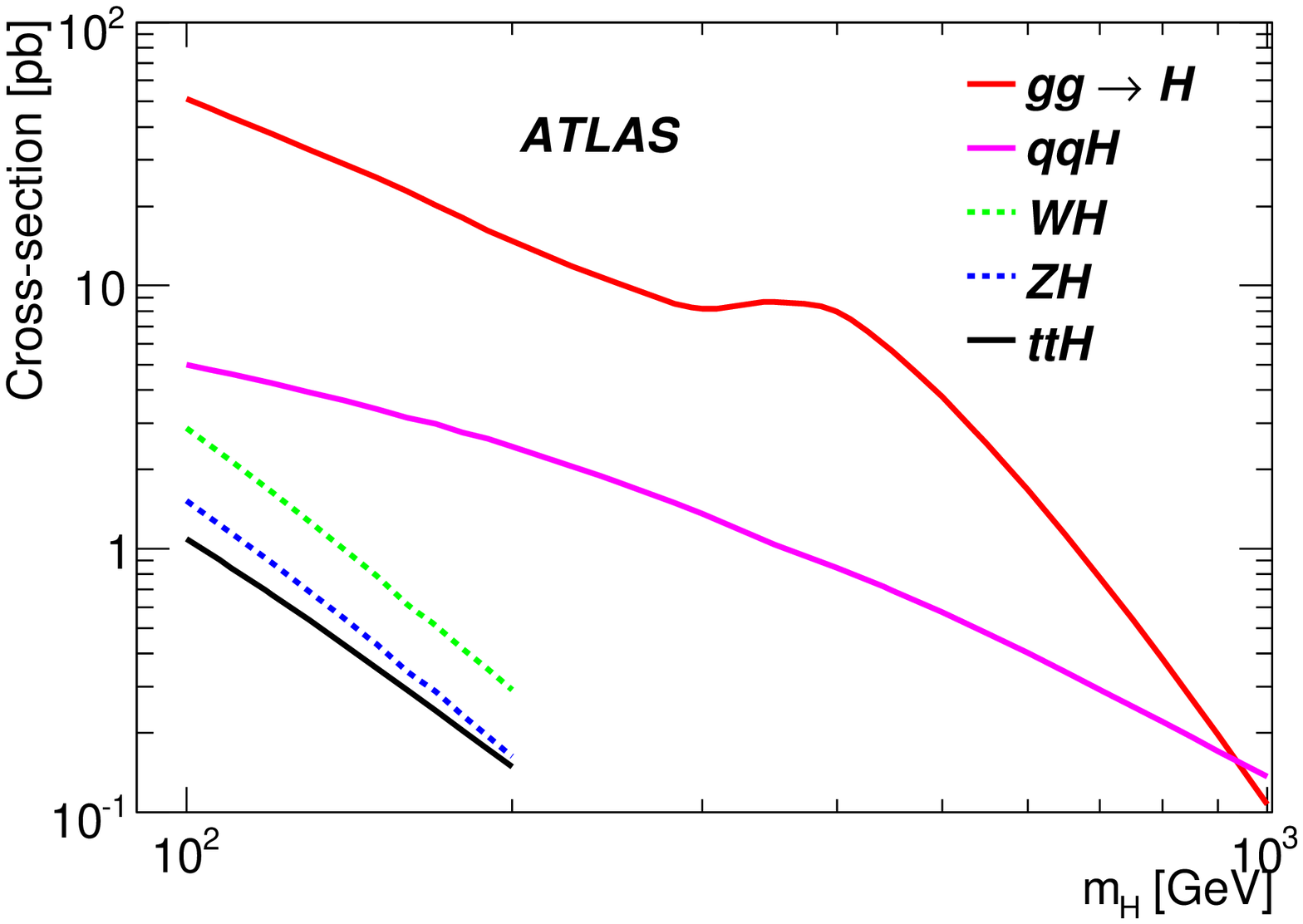,width=5.0cm, height=5.0cm}
  \epsfig{file=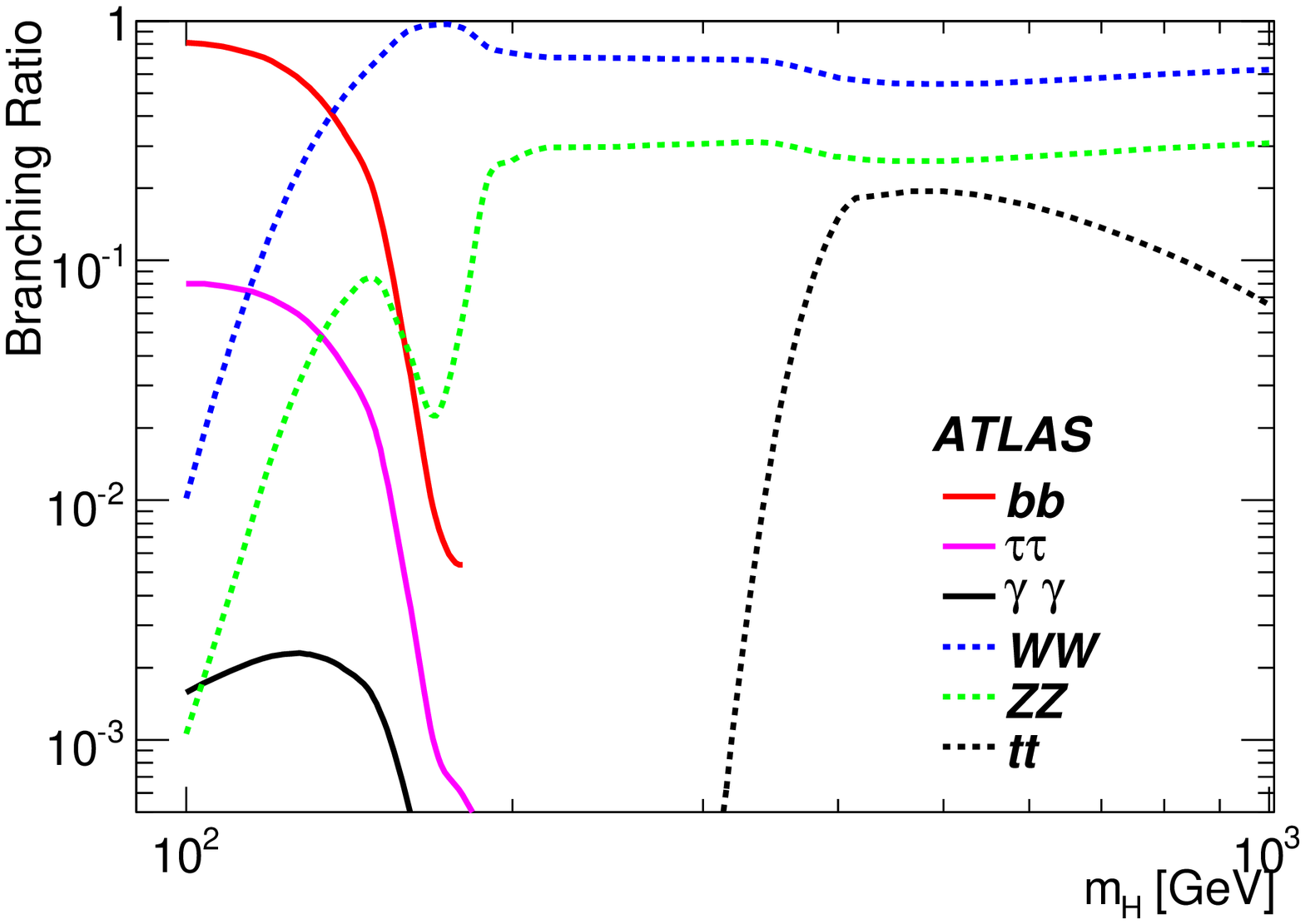,width=5.0cm, height=5.0cm}\\
}
\end{center}
\caption{\em Right: cross-sections for the five production channels of the Standard Model Higgs 
boson at the LHC at 14 TeV. Left: branching ratio for the relevant decay modes of the Standard 
Model Higgs boson as a function of its mass. From Ref.~\cite{CERN-OPEN-2008-020}. 
\label{FS2.1}}
\end{Fighere}

\section{Prospects in the Standard Model Higgs boson searches}

 The Standard Model Higgs boson will be searched for at the LHC in various decay channels. Over the past 
ten years, strategies for signal observability and background rejection methods have been established 
in many studies~\cite{ATLASTDR}. 

In the following I will discuss four of the most important decay channels, for which studies have been recently 
revisited with the ATLAS experiment and which are included in the most recent combination results.

\subsection{ The $H \to \gamma \gamma$ decays}

The decay  $H \to \gamma \gamma$ is a rare decay mode, which is detectable only in the limited Higgs 
boson mass region  100-150 GeV. Excellent energy and angular resolution are required to observe
the narrow mass peak above the irreducible prompt $\gamma \gamma$ continuum and the reducible background 
resulting from direct photon production or from two-jet production via QCD jets processes. 
In the recent evaluation different topologies have been studied, i.e. only inclusive analysis 
requiring photon pair in the final state or in association with one or two high $p_T$ jets. 
In addition to the diphoton invariant mass, other discriminating variables are incorporated into the 
analysis and combined by means of an unbinned maximum-likelihood fit. Photon reconstruction properties 
and the event topology are used to separate the sample into categories that are fit simultaneously.
The inclusive analysis leads to the highest expected  signal but also background rates.  
The expected cross-sections (in fb) are given in Table~\ref{TS2.1}. The relative contribution from events 
with at least one fake photon constitutes 39\% of the total background.
The theoretical uncertainties of the predictions for prompt single and double photon 
production used in the inclusive analysis have been evaluated and are summarised in Table~\ref{TS2.2}.
Diphoton invariant mass spectrum after the application of cuts of the inclusive analysis is shown
in Fig.~\ref{FS3.1} (left). The hatched histograms present contribution from events with one or two 
fake photons. 
Selecting more exclusive topologies like $H+1j$, $H+2j$ leads to better signal-to-background ratio but
much fewer signal events.

\begin{Tabhere} 
\newcommand{\lstrut}{{$\strut\atop\strut$}}
  \caption {\em Expected cross-sections (in fb) for different signal ($m_H$=120 GeV) and background processes
within a mass window of $m_{\gamma \gamma} \pm 1.4$ of the mass resolution (1.46 $\pm$ 0.01) GeV in the 
no-pileup case and after selection for inclusive analysis. From Ref.~\cite{CERN-OPEN-2008-020}.
\label{TS2.1}}
\vspace{2mm}   
\begin{center}
\begin{tabular}{|cc|cc|} \hline
Signal Process & Cross-section (fb) & Background Process & Cross-section (fb) \\
\hline 
$gg \to H$     &      21            & $\gamma \gamma$         &   562  \\
$VBF H$        &      2.7           & Reducible $\gamma j$    &   318  \\
$t \bar t H$   &      0.35          & Reducible $j j$         &    49  \\
$ V H$         &      1.3           & $ Z \to e^{+} e^{-}$    &    18  \\
\hline 
\end{tabular}
\end{center} 
\end{Tabhere}

\begin{Tabhere} 
\newcommand{\lstrut}{{$\strut\atop\strut$}}
  \caption {\em Summary of the relative systematic uncertainties on the $\gamma \gamma$ and $\gamma j$ processes.
 From Ref.~\cite{CERN-OPEN-2008-020}.
\label{TS2.2}}
\vspace{2mm}   
\begin{center}
\begin{tabular}{|ccc|} \hline
Potential sources & $\gamma \gamma$ & $\gamma j$ \\
\hline
Scale dependence     &  14\% & 20\%  \\
Fragmentation        &  5\%  & 1\%   \\
PDF                  &  6\%  & 7\%   \\
\hline
Total                &  16\% & 21\%  \\
\hline 
\end{tabular}
\end{center} 
\end{Tabhere}

\begin{Fighere}
\begin{center}
{
  \epsfig{file=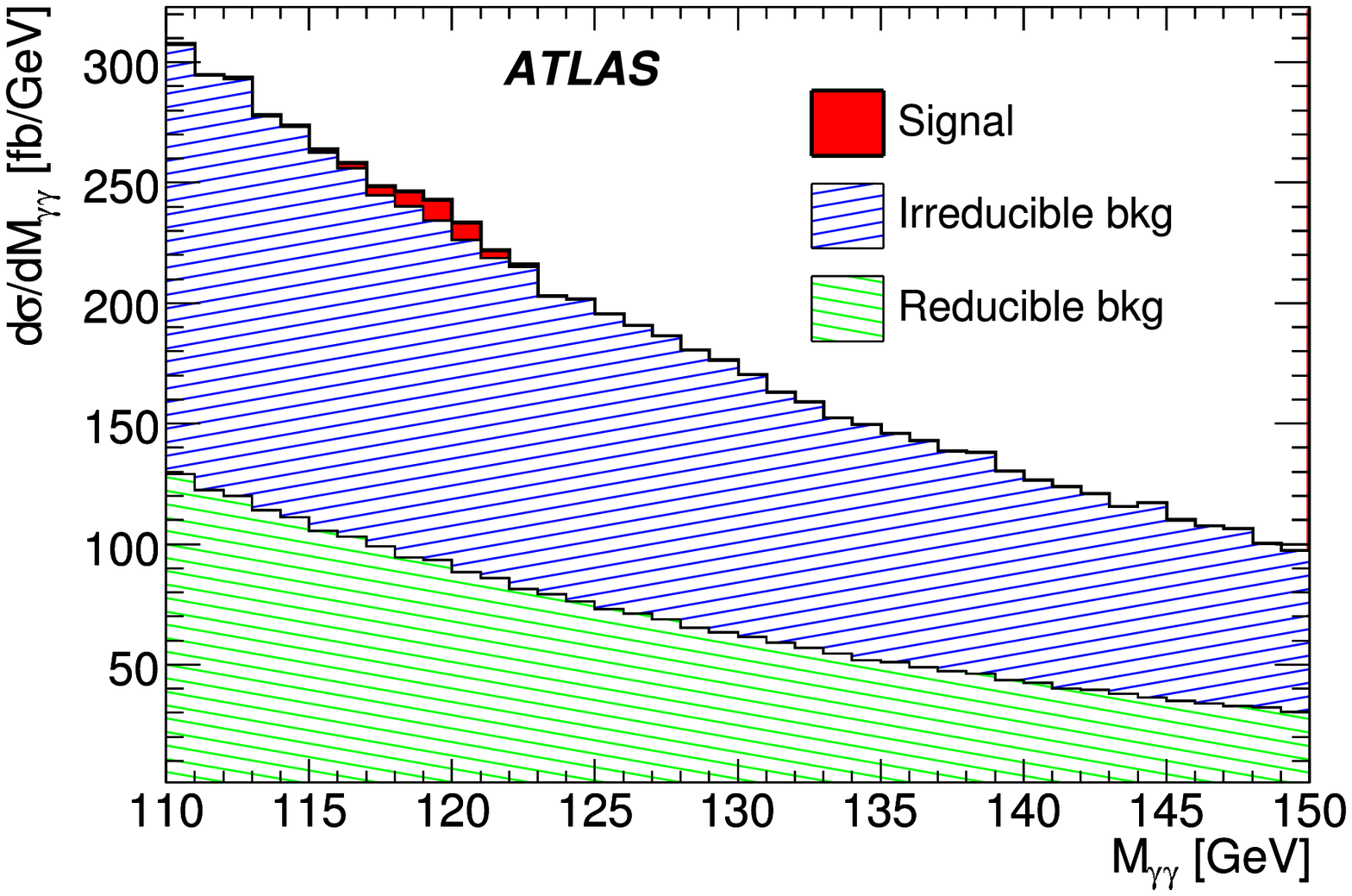,width=5.0cm, height=5.0cm}
  \epsfig{file=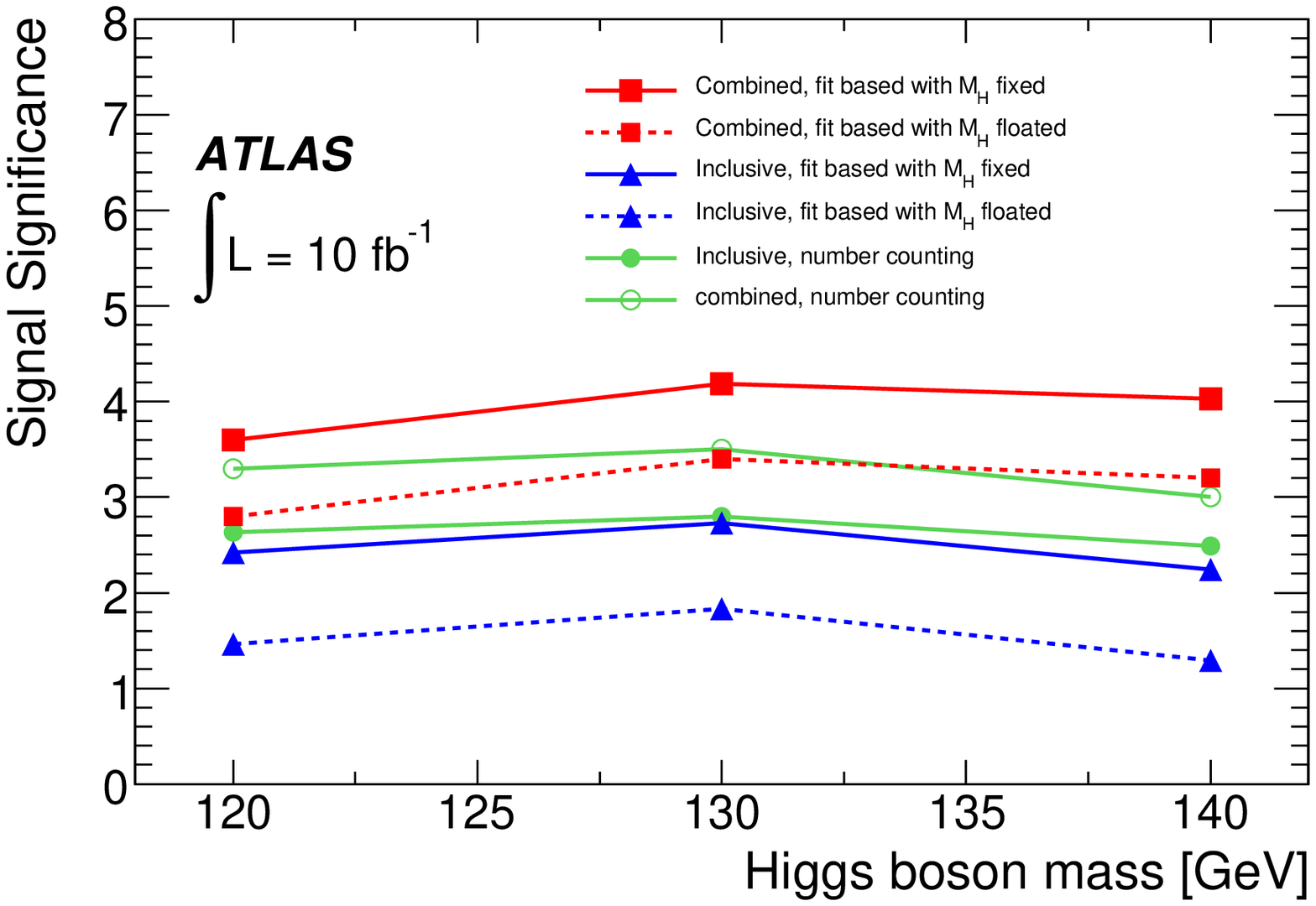,width=5.0cm, height=5.0cm}\\
}
\end{center}
\caption{\em Left:  diphoton invariant mass spectrum after the application of cuts of the inclusive analysis. 
Right: expected signal significance, as a function of the Higgs mass, for the $H \to \gamma \gamma$ 
decay assuming an integrated luminosity of 10 fb$^{-1}$. 
 From Ref.~\cite{CERN-OPEN-2008-020}. 
\label{FS3.1}}
\end{Fighere}

The expected signal significance for the Higgs boson using the $H \to \gamma \gamma$ decay for 10 fb$^{-1}$
of the integrated luminosity as a function of mass is shown in Fig~\ref{FS3.1}. The sensitivity of the
inclusive analysis based on the events counting and the sensitivity when the Higgs boson plus jet is required
are shown separately. In addition, shown is also the sensitivity obtained using more elaborate techniques like 
one dimensional fits with a fixed and floating Higgs boson mass, respectively.
A signal significance  based on events counting of 2.6 (4.6) can be reached with an integrated luminosity 
of 10 (30) fb$^{-1}$ for a Higgs mass of 120 GeV in the case of inclusive analysis. 
The addition of search in association with jets enhances the event counting signal significance to 3.3 (5.7) 
with 10 (30) fb$^{-1}$. 
The sensitivity can be further enhanced by means of unbinned maximum-likelihood fit dividing samples into 
categories, depending on the event topology and exploiting a number of discriminating variables. An increase 
up to 3.6 (2.8) at 10 fb$^{-1}$  integrated luminosity in case of fixed (floating) mass fit can be expected.

\subsection{ The $H \to ZZ^{(*)} \to 4 \ell $ and $H \to ZZ \to 4 \ell $ decays}

The decay channel  $H \to ZZ^{(*)} \to 4 \ell $ provides a rather clean signature in the mass 
range  120 GeV $< m_{H} <$ 2 $m_Z$ and is the most reliable one for the discovery in the range 
2 $m_Z$ $< m_{H} <$ 700 GeV. In addition to the irreducible backgrounds from $ZZ^*$ and $Z\gamma^*$
production, there are large reducible backgrounds from $t \bar t$ and $Zb \bar b$ production.
It has been shown that with expected performance of the detector \cite{ATLASDetPaper} the reducible background 
can be suppressed well below the irreducible  $ZZ^*/\gamma^* \to 4 \ell$, calorimeter and track isolation 
of leptons together with impact parameter measurements can be used to achieve necessary background rejection  
against QCD jets and non-isolated leptons from semileptonic decays of heavy flavour quarks. 
The NLO cross-sections after the full event selection, are shown in Table~\ref{TS2.3} for the three decay 
channels combined .

\begin{Tabhere} 
\newcommand{\lstrut}{{$\strut\atop\strut$}}
  \caption {\em Expected cross-sections (in fb) for signal and backgrounds after full selection within 
the mass window $m_{H} \pm \sigma_{m_{H}}$, where $\sigma_{m_{H}}$ is the experimental 4-lepton 
mass resolution, $\sigma_{m_{H}} = 1.4\% - 1.8 \%~m_{H}$ for the mass values shown here.
The expected significance is given for an integrated luminosity of 30 $fb^{-1}$. Systematic errors are not 
yet taken into account in significance calculation. The $t \bar t$ background is assumed
not to contribute to significance. 
\label{TS2.3}}
\vspace{2mm}   
\begin{center}
\begin{tabular}{|c| ccc|} \hline 
Mass  &  130 GeV & 150 GeV & 180 GeV \\
\hline 
Signal             &  0.816 &  1.94  & 1.32   \\
$ZZ^*/\gamma^*$    &  0.150 &  0.151 & 0.938  \\
$Z b \bar b$       &  0.047 &  0.021 & 0.013  \\
 $ t \bar t$       &  $<$ 0.04 & $<$ 0.04 &  $<$ 0.04  \\
\hline
Significance (30 fb$^{-1}$) & 7.1 & 14.2& 6.2 \\
\hline 
\end{tabular}
\end{center} 
\end{Tabhere}

\begin{Fighere}
\begin{center}
{
  \epsfig{file=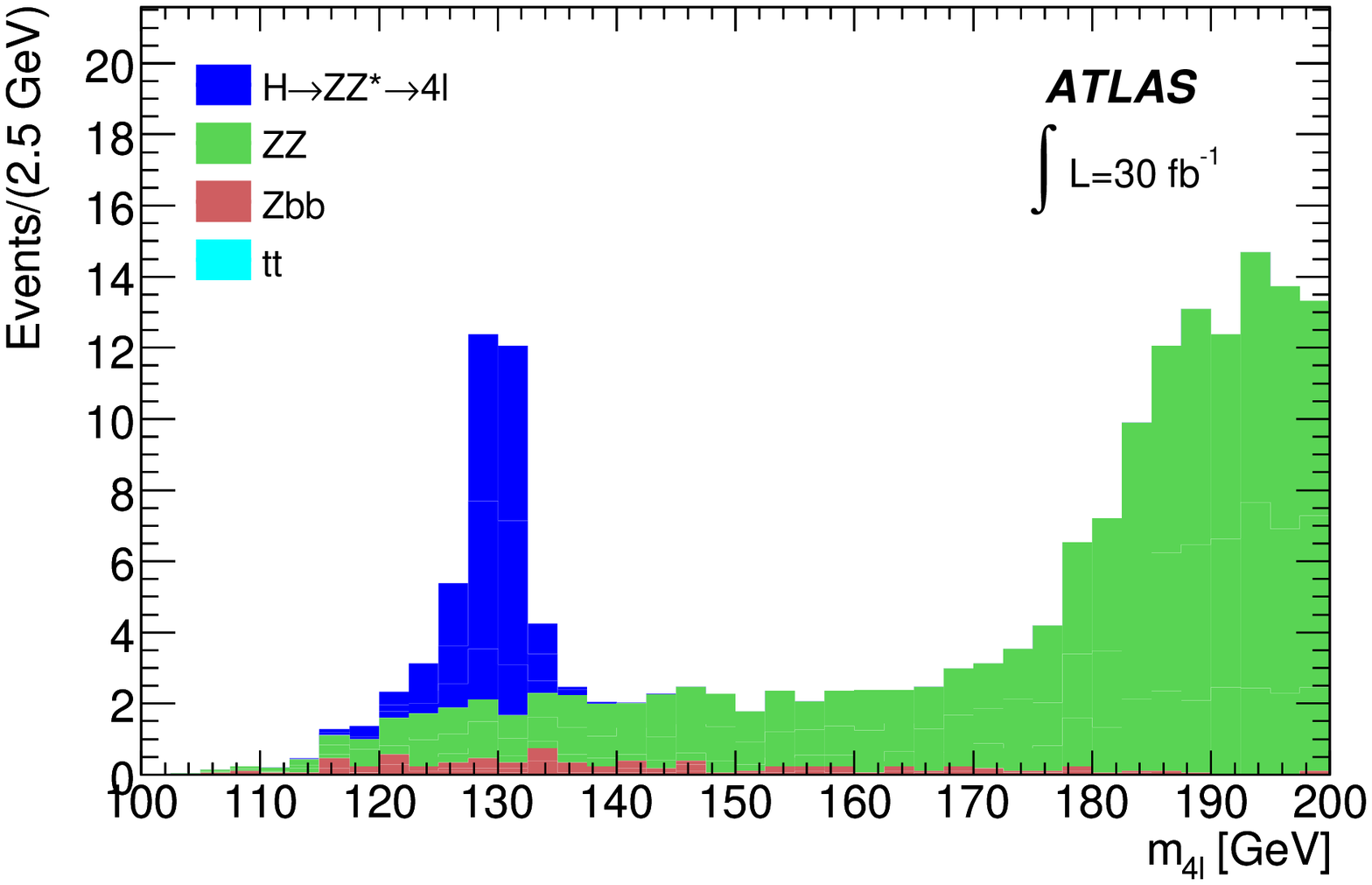,width=5.0cm, height=5.0cm}
  \epsfig{file=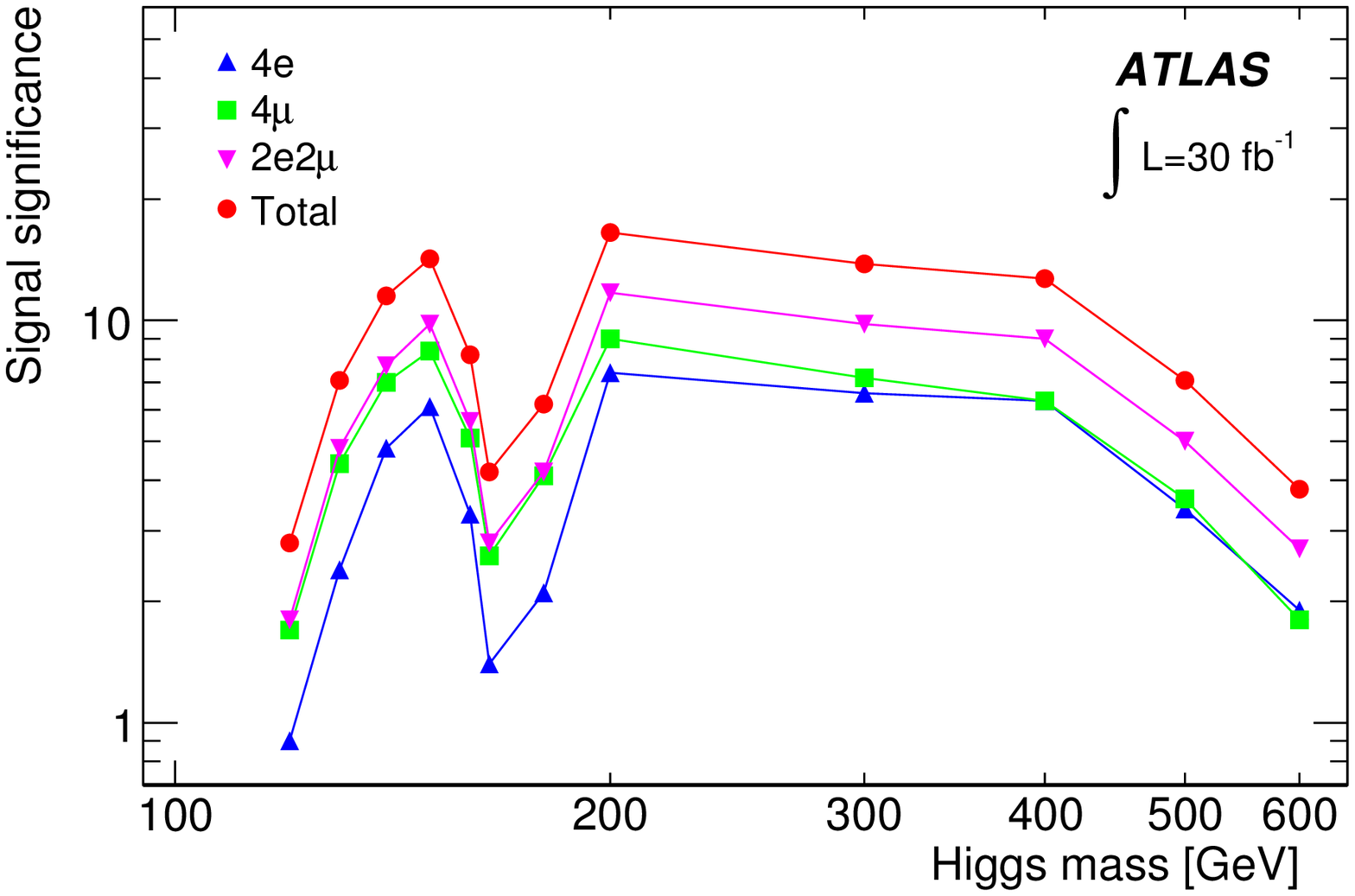,width=5.0cm, height=5.0cm}\\
}
\end{center}
\caption{\em Reconstructed 4-lepton mass for signal and background processes, in the case of
a 130 GeV Higgs boson, normalized to an integrated luminosity of 30 fb$^{-1}$.
Expected signal significances are computed using Poisson statistics, for each of the three decay channels, 
and their combination (systematic uncertainties not included). From Ref.~\cite{CERN-OPEN-2008-020}. 
\label{FS3.2}}
\end{Fighere}

The reconstructed  4-lepton mass for signal and background processes, in the case of
a 130 GeV Higgs boson, normalized to the luminosity of 30~fb$^{-1}$ is shown in Fig.~\ref{FS3.2} (left).
The expected signal significances computed using Poisson statistics, for each of the three decay channels, 
and their combination are shown in Fig.~\ref{FS3.2} (right) without including systematic uncertainties.
The significance is slightly reduced if the systematic uncertainties are included;
in this case the significance is evaluated using the profile likelihood ratio method~\cite{CERN-OPEN-2008-020}.
For an integrated luminosity of 30 fb$^{-1}$, the discovery of the 
Higgs boson in the $4 \ell$ channel alone in a mass range 130-500 GeV will be possible, with the 
exception of the region around 160 GeV of the $WW$ turn on.
The  $H \to ZZ \to 4 \ell $ channel is highly sensitive in the high mass region, 200-400 GeV, and in the 
150 GeV region where the Higgs boson should be discovered with an integrated luminosity of 5 fb$^{-1}$.

\subsection{ The $qqH \to qq \tau \tau $  decays}

A significant discovery potential in the low mass region can be expected
when searching for the decay into pair of $\tau$ leptons and of Higgs bosons produced in association 
with two jets (mainly by the Vector Boson Fusion  process). The analysis requires excellent 
performance for every detector subsystem: the presence of the $\tau$ decays implies final states 
with electrons, muons, hadronic $\tau$ decays and missing transverse momentum, 
while the Vector Boson Fusion production process introduces jets that tend to be quite forward
in the detector. The sensitivity is estimated based on the combination of the lepton-lepton ($\ell \ell$)
and lepton-hadron ($\ell h$) topology for $\tau$-pair decays;
the performance in hadron-hadron ($hh$) channel has been also addressed in most 
recent analyses~\cite{CERN-OPEN-2008-020}.

\begin{Fighere}
\begin{center}
{
  \epsfig{file=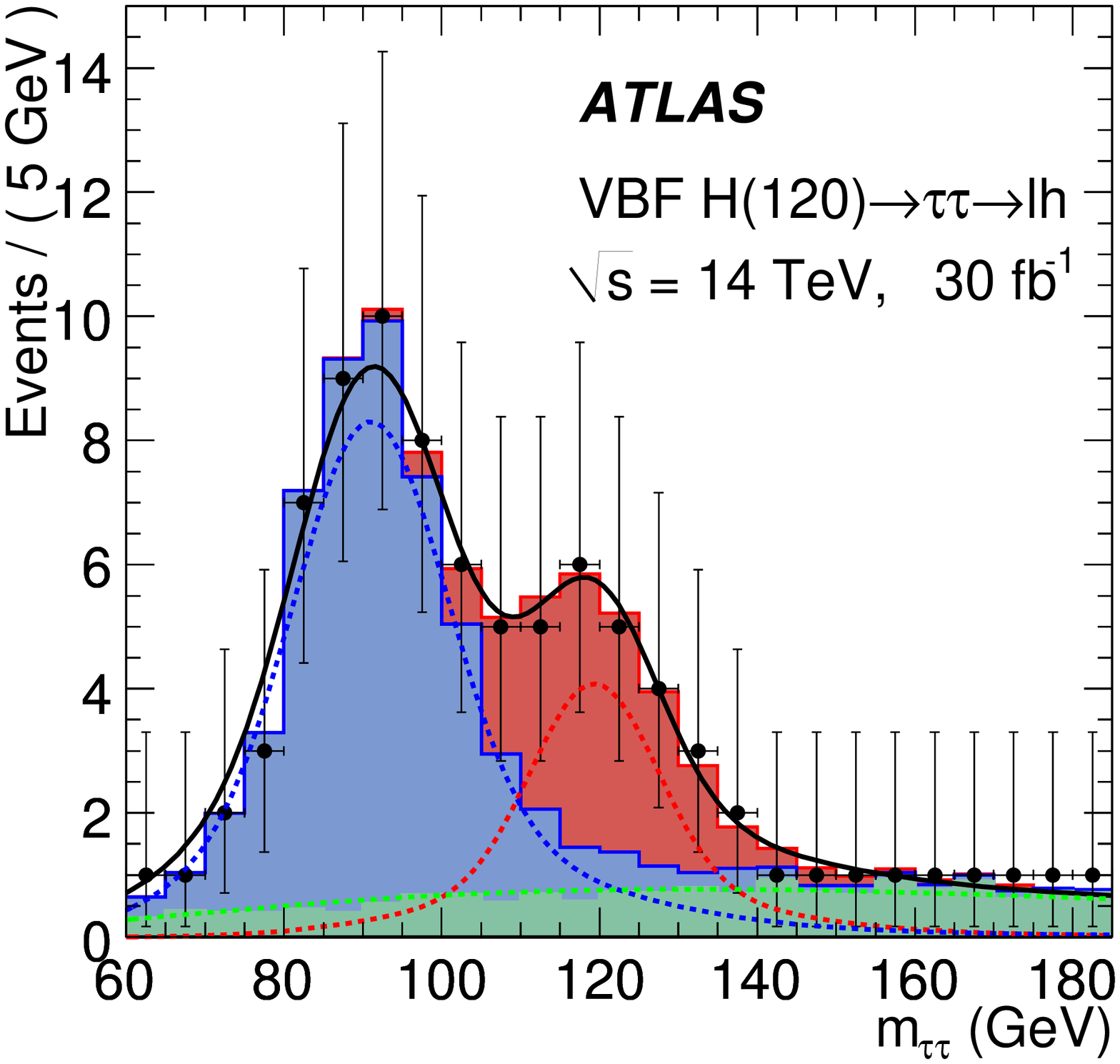,width=5.0cm, height=4.5cm}
  \epsfig{file=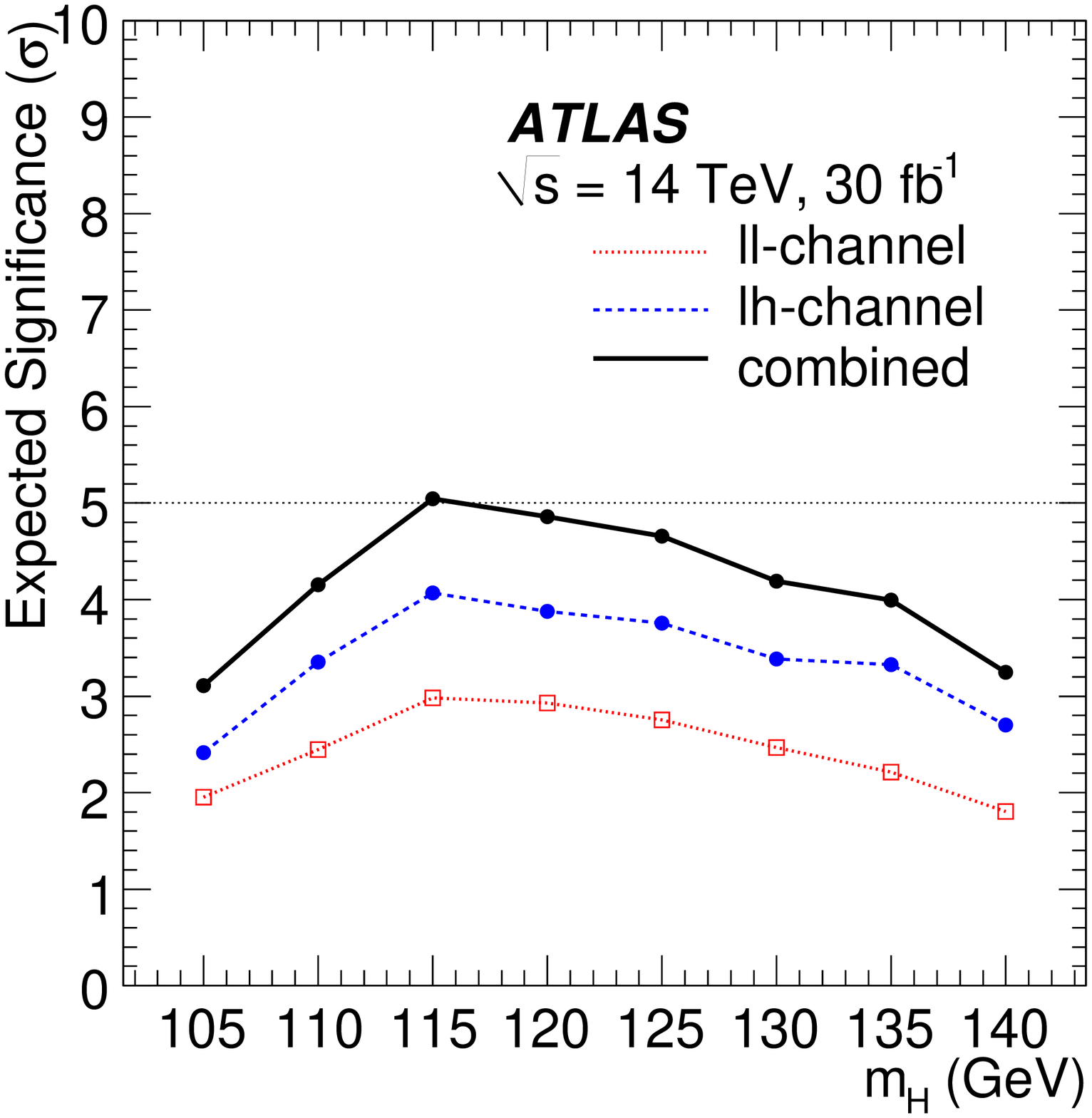,width=5.0cm, height=4.5cm}\\
}
\end{center}
\caption{\em Example fits to a data sample with the signal-plus-background for the lh- channel
 at $m_{H} =$ 120 GeV with 30 fb$^{-1}$ of data.
Expected signal significance for several masses based on fitting the $m_{\tau \tau}$ spectrum. 
Background uncertainties are incorporated by utilizing the profile likelihood ratio method. 
These results do not include the impact of pileup. From Ref.~\cite{CERN-OPEN-2008-020}. 
\label{FS3.3}}
\end{Fighere}

The signal events are produced with significant transverse momenta, so the 
$\tau$ from the decay are boosted which causes their decay products being almost collinear in the 
laboratory frame. The $\tau \tau$ invariant mass can be therefore reconstructed in the collinear 
approximation, i.e. assuming that the $\tau$ direction is given by their visible decay products 
(charged lepton or hadrons). The mass resolution is about 10 GeV, leading to approximately 3.5\% 
precision on the mass measurement with 30 fb$^{-1}$ of data.  
Example fits to a data sample with the signal-plus-background 
for the lh- channel at $m_{H}$ = 120 GeV with 30 fb$^{-1}$ of data is shown in Fig.~\ref{FS3.3} (left). 

In the recent analysis particular emphasis has been placed on data-driven background estimation 
strategies and the associated uncertainties in normalisation and shape. 
Expected  signal significance for several masses based on fitting the $m_{\tau \tau}$ spectrum is shown 
in  Fig.~\ref{FS3.3} (right).
The results obtained neglecting the pileup effects indicate that a  $\sim$ 5$\sigma$ significance 
can be achieved for the Higgs boson mass in the range 115-125~GeV
after collecting 30 fb$^{-1}$ of data and combining the $\ell \ell$ and $\ell h$ channels. 
The effects induced by the event pile-up has not been fully addressed yet.

\subsection{ The $gg \to H \to WW$ and $qqH \to qq WW $  decays}

The prospects for the Higgs searches mass range below 200 GeV through the channel
$H+0j$ and the VBF channel $ H+2j$  with $H \to WW \to e \nu \mu \nu$
have been reviewed, with an emphasis on evaluating methods to estimate backgrounds using 
control samples in data. In the studies of the former one must
consider background processes that yield two leptons and significant missing transverse energy 
in the final state; in the latter, the final state also includes two hard jets 
(from the struck quarks) which tend to be well-separated in pseudorapidity. 
In this channel it is not possible to reconstruct the Higgs boson mass peak; instead, an excess
of events above the expected backgrounds can be observed and used to establish the presence 
of the Higgs boson signal, provided the background level can be safely controlled.
 Usually, the transverse mass computed from the leptons and the missing 
transverse momentum is used to discriminate between signal and background. In addition, its shape
provides additional sensitivity to the true Higgs boson mass. 

Signal discrimination from the background 
relies on a complicated topological selection. The  $H+0j$ analysis requires two opposite-sign 
isolated leptons, large missing transverse energy, vetoes additional jets and exploits
differences from spin correlations effects between signal and backgrounds on the transverse 
opening angle between leptons.

\begin{Tabhere} 
\newcommand{\lstrut}{{$\strut\atop\strut$}}
  \caption {\em Expected cross-sections (in fb) after selection for a cut based analysis with 
a test mass $m_H = 170$ GeV in the $H+0j, H \to WW \to e \nu \mu \nu$ channel. 
The $W+jets$ and $b \bar b$ backgrounds are omitted for this table.
\label{TS2.4}}
\vspace{2mm}   
\begin{center}
\begin{tabular}{|c| cccc|} \hline 
Region             &  Signal           & $t \bar t$       & $WW$             & $Z \to \tau \tau$  \\
\hline 
Signal-like        &  28.65 $\pm$ 0.80 & 1.14 $\pm$ 1.14  & 29.35 $\pm$ 1.59 & $<$1.74          \\
Control            &  1.47 $\pm$ 0.27  & 5.71 $\pm$ 2.55  & 61.13 $\pm$ 2.33 & 4.06 $\pm$ 1.53  \\
b-tagged           &    0              & 6.85 $\pm$ 2.80  & 0.11  $\pm$ 0.09 & 1.16 $\pm$ 0.82  \\
\hline 
\end{tabular}
\end{center} 
\end{Tabhere}

The lepton selection (identification and isolation) criteria are 
more stringent than in discussed latter $H+2j$ analysis, since  $H+0j$ signature has fewer handles 
to suppress large reducible backgrounds from $W+jets$ and $b \bar b$ processes.
In Table~\ref{TS2.4} shown is an example of separating signal 
and background regions with different kinematical selections which can be then used for in-situ 
normalisation and cross-check on the background predictions.

Fig.~\ref{FS3.4} (left) shows the distribution of the 
transverse opening angle $\Delta \phi^{\ell \ell}$  of the two leptons after preselection cuts
for signal and $WW$ background. Fig.~\ref{FS3.4} (right) shows the  distribution of the transverse
 mass $m_T$ for events with 
$\Delta \phi^{\ell \ell} <$ 1.575 and $p_T^{WW} >$ 20~GeV, 
in a fitted toy Monte Carlo outcome containing the Standard Model Higgs boson with 
$m_H$ = 170 GeV, after 10~fb$^{-1}$ of the integrated luminosity.

\begin{Fighere}
\begin{center}
{
  \epsfig{file=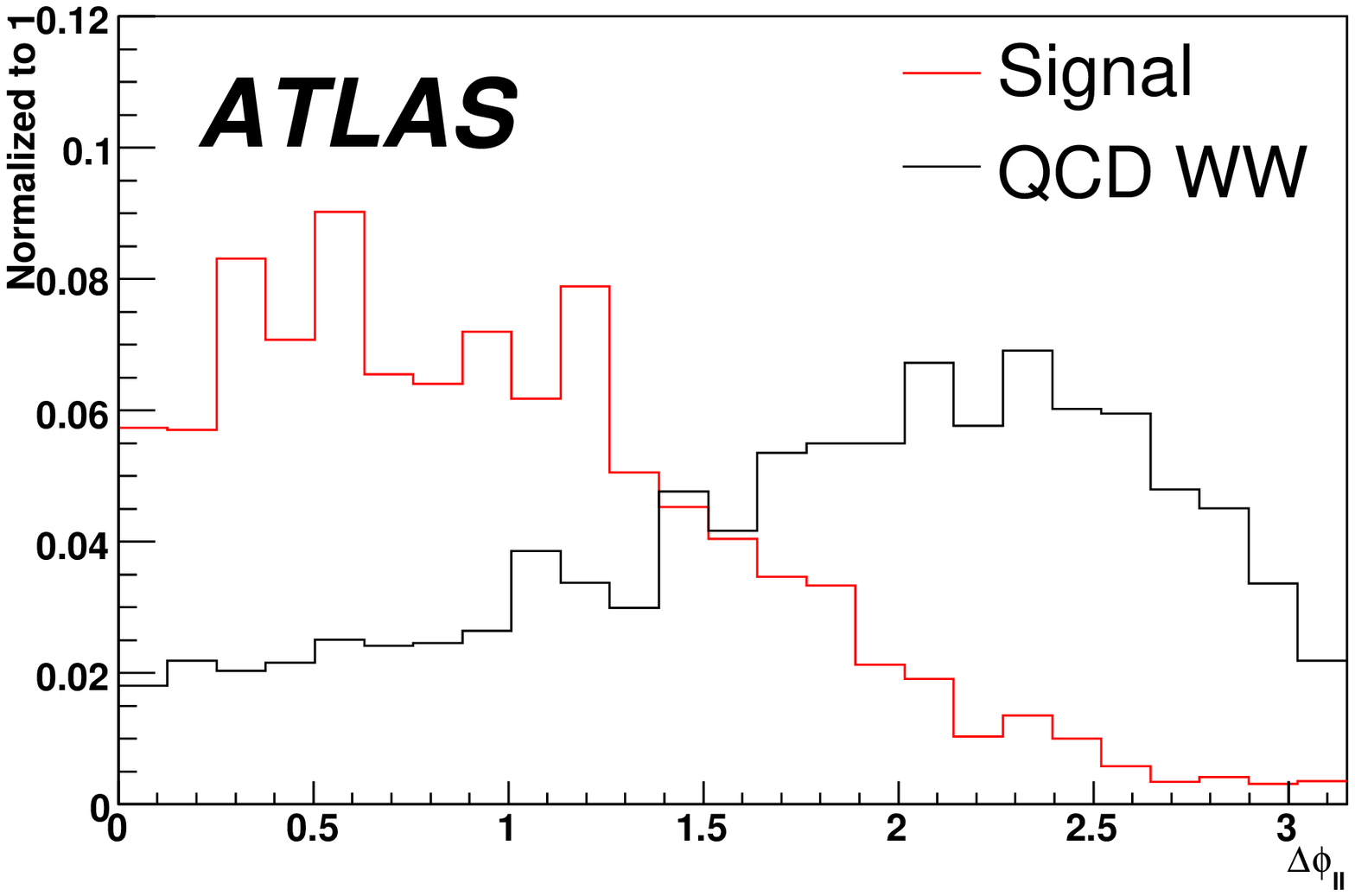,width=5.0cm, height=4.5cm}
  \epsfig{file=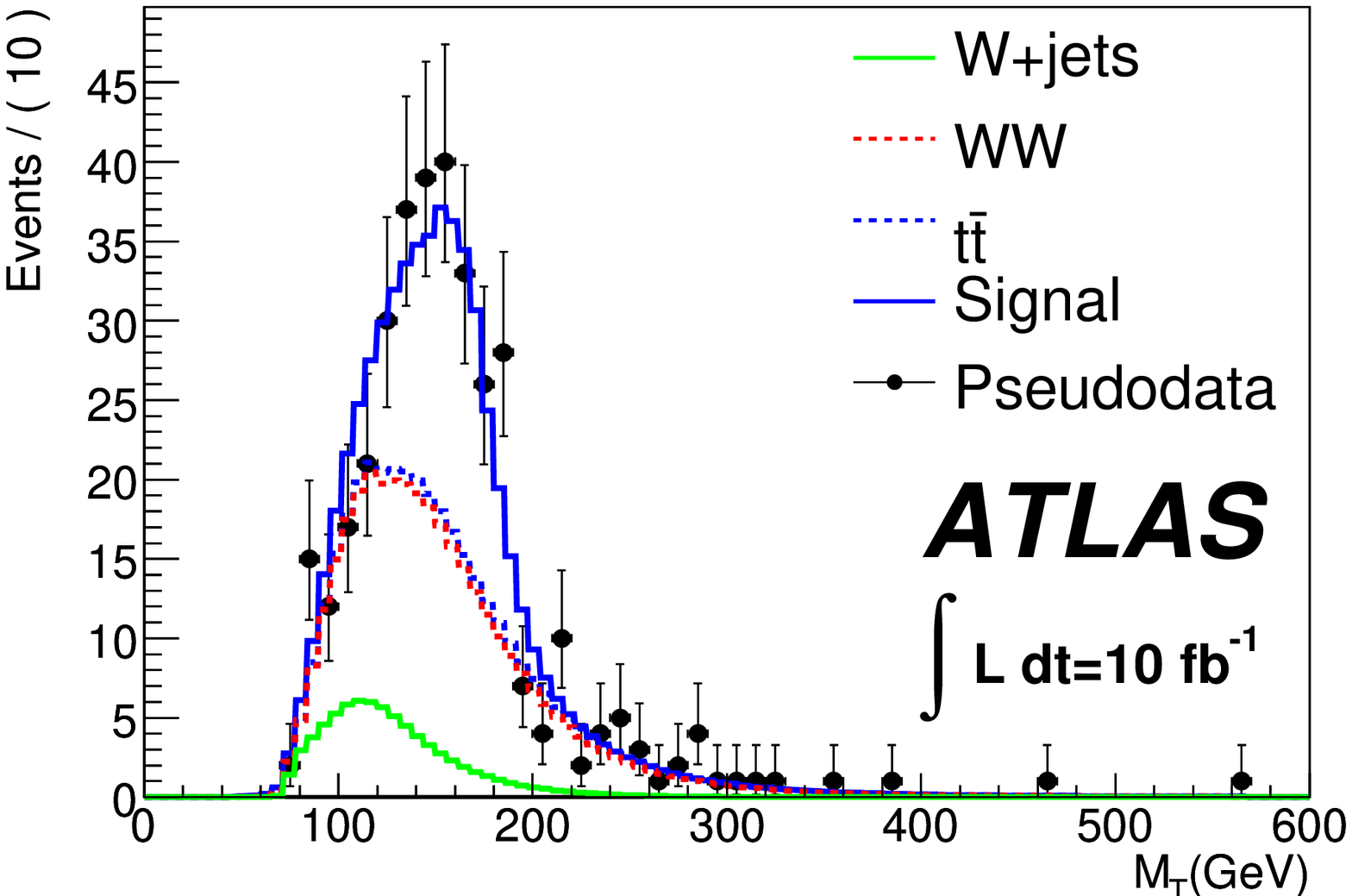,width=5.0cm, height=4.5cm}\\
}
\end{center}
\caption{\em
Left: transverse opening angle $\Delta \phi^{\ell \ell}$  of the two leptons after preselection cuts.
Right: transverse mass $m_T$ for events with $\Delta \phi^{\ell \ell} <$ 1.575 and 
$p_T^{WW} >$ 20~GeV (see text). 
From Ref.~\cite{CERN-OPEN-2008-020}. 
\label{FS3.4}}
\end{Fighere}

\begin{Fighere}
\begin{center}
{
  \epsfig{file=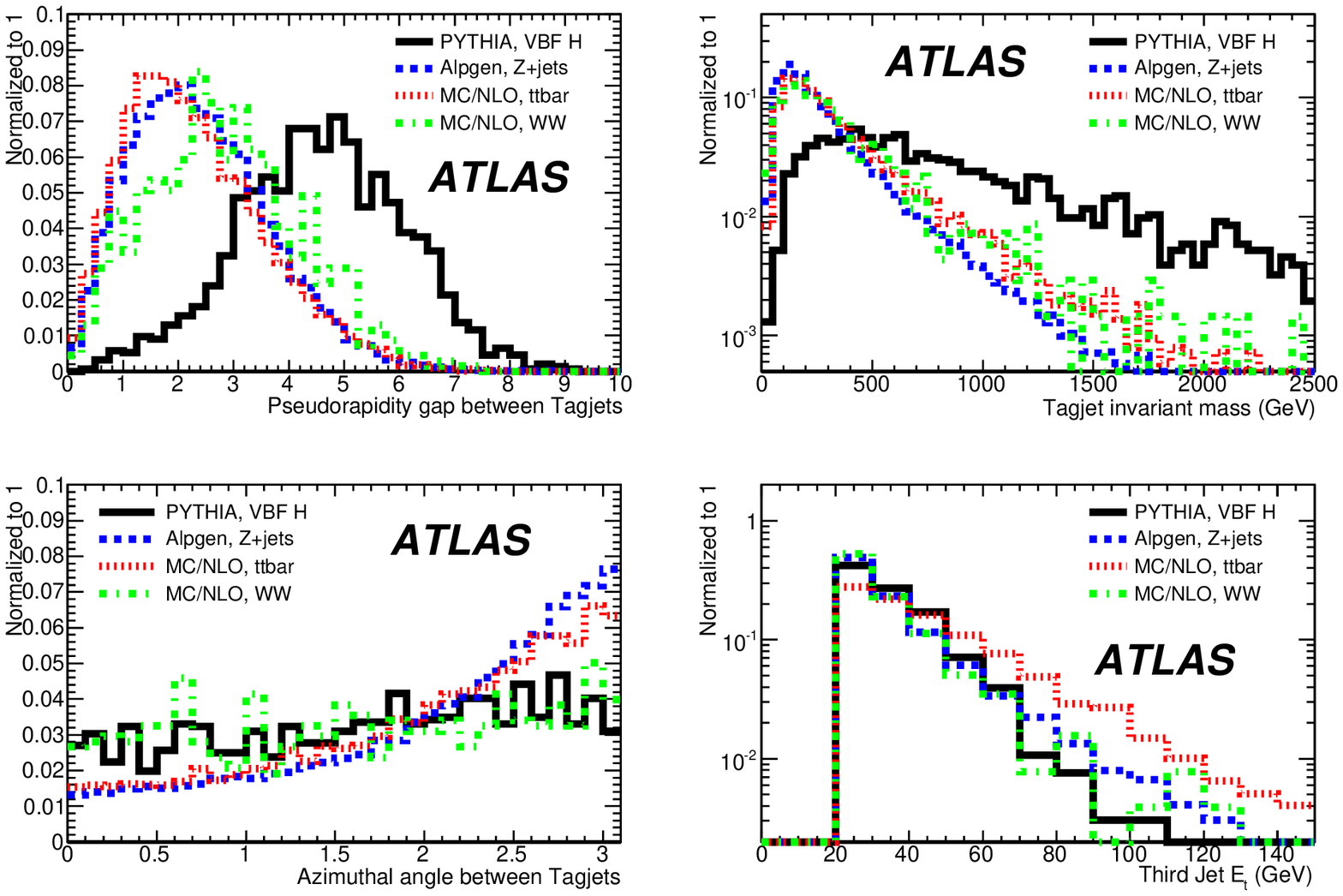,width=10.0cm, height=7.0cm}
}
\end{center}
\caption{\em
Pseudorapidity gap between tag jets (left-top), invariant mass distribution of tag jets (right-top),
azimuthal angle gap between jets (left-bottom) and transverse energy of the third one (left-right).
From Ref.~\cite{CERN-OPEN-2008-020}. 
\label{FS3.5}}
\end{Fighere}
\vspace{2mm}

\begin{Fighere}
\begin{center}
{
  \epsfig{file=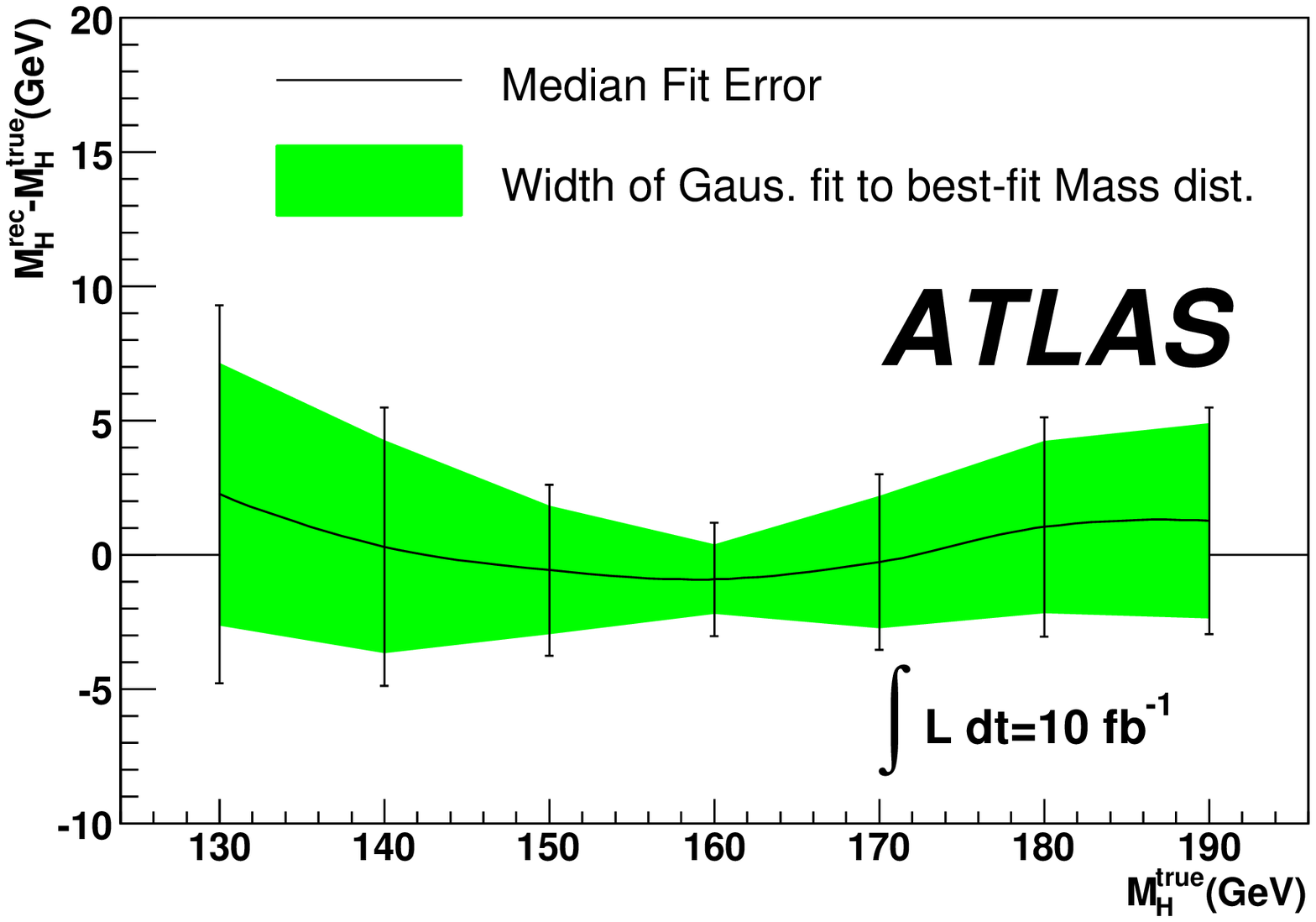,width=5.0cm, height=4.5cm}
  \epsfig{file=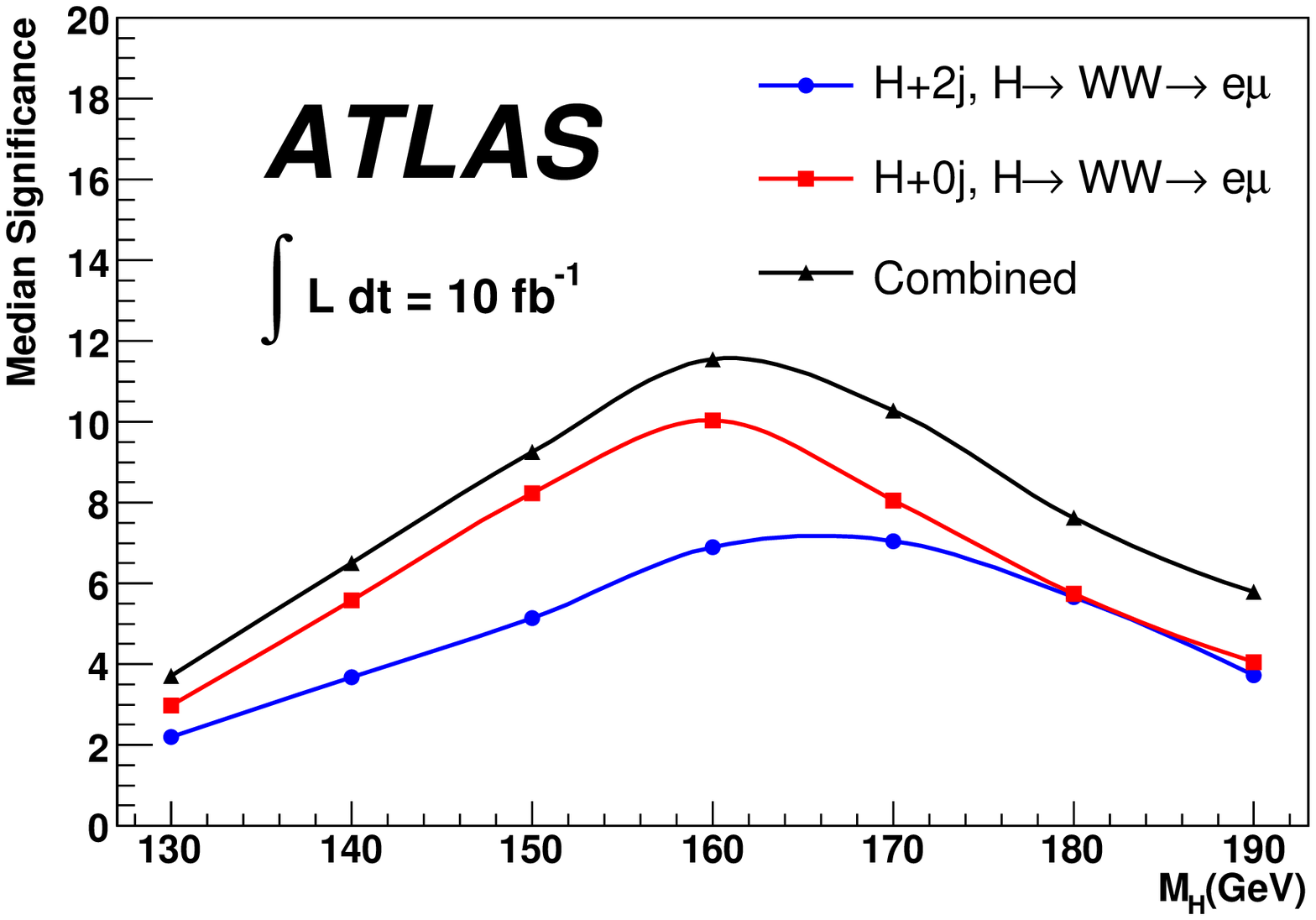,width=5.0cm, height=4.5cm}\\
}
\end{center}
\caption{\em 
Left: the linearity of the mass determination for the combined fit of $H + 0/2j$, $H \to WW \to e \nu \mu \nu$.
Right: the expected significance at 10 $fb^{-1}$.
From Ref.~\cite{CERN-OPEN-2008-020}. 
\label{FS3.6}}
\end{Fighere}

The $H+0j$ channel is very promising for the Higgs boson masses around the $WW$ threshold. 
An in-situ background normalisation technique has been prepared and it has been shown that 
the uncertainties can be controlled, and that background can be normalised with a two-dimensional
fit in the transverse mass and the $p_T$ of the $WW$ system. 
With 10 fb$^{-1}$ one would 
expect to be able to reach a 5$\sigma$ discovery with the $H \to WW \to e \nu \mu \nu$ channel alone 
if there is the Standard Model Higgs boson with a mass between 140 and 180 GeV.

The $H+2j$ analysis explores, in addition, characteristic kinematical 
features of the accompanying jets (tag jets) from VBF production: pseudorapidity gap between jets, invariant 
mass of the jet-system and vetoes additional jet activity in the gap region. In Fig.~\ref{FS3.5}
shown are distributions essential for discriminating signal and background events. A requirement
of $\eta_1 \eta_2 <$ 0 and jet $E_T > 20$ GeV is used. Distributions are shown for $m_H = 170$~GeV,
but the dependence of these variables on the Higgs boson mass is rather weak.
This channel has smaller event rate than the  $H+0j$ channel
but leads to a similar significance. 

With  10 fb$^{-1}$ integrated luminosity, one should 
expect to be able to reach a $5\sigma$ discovery in the $H \to WW \to e \nu \mu \nu$ channel 
alone if there is the Standard Model Higgs boson with a mass between 155 and 180~GeV.  
Combining the two channels here discussed one would expect to be able to
discover the Standard Model Higgs boson in the $H \to WW$ decay mode in the mass
range 135-190 GeV.
Fig.~\ref{FS3.6} shows the linearity of the mass 
determination and the expected significance of a combined fit of the two topologies as a function
of a true Higgs boson mass. The green band represents the width of the gaussian fit to the region 
around the peak of the best-fit mass distribution and the error bars show the  median fit error. 
With 10 fb$^{-1}$ integrated luminosity one would expect to be able to measure Higgs boson
mass with a precision of about 7 GeV for a true mass 130 GeV or about 2 GeV for a 
true mass  160 GeV.

\subsection{ Statistical combination of Standard Model channels}

The Higgs boson searches will exploit a number of statistically independent decay channels.  
All of the information will be combined to provide a single estimate of the significance
of a discovery or exclusion limits on the Higgs production. 
The approach taken in most recent studies \cite{CERN-OPEN-2008-020}
is based on frequentist statistical methods, where effects of systematic uncertainties 
are incorporated by the use of the profile likelihood ratio. 
The profile likelihood ratio treats systematic errors by associating the uncertainties with
adjustable (nuisance) parameters and adjusting those to maximize the likelihood. 

For different hypothesised Standard Model Higgs boson mass, the  signal significances expected
are presented  assuming the Standard Model Higgs production rate, as well as the expected
upper limits on the Higgs production cross-section, under the hypothesis of no Higgs signal.
The studies have exploited a series of useful approximations that allow one to determine the 
median discovery and exclusion sensitivities from a combined fit.
The validation studies indicate that 
the approximations used should be reasonably accurate or lead to conservative limits for 
integrated luminosity over 2~fb$^{-1}$. For earlier stages of the experiment it 
is expected that one will need to rely on Monte Carlo methods, which should be feasible
for exclusion limits at the 95\% confidence level.

\begin{Fighere}
\begin{center}
{
  \epsfig{file=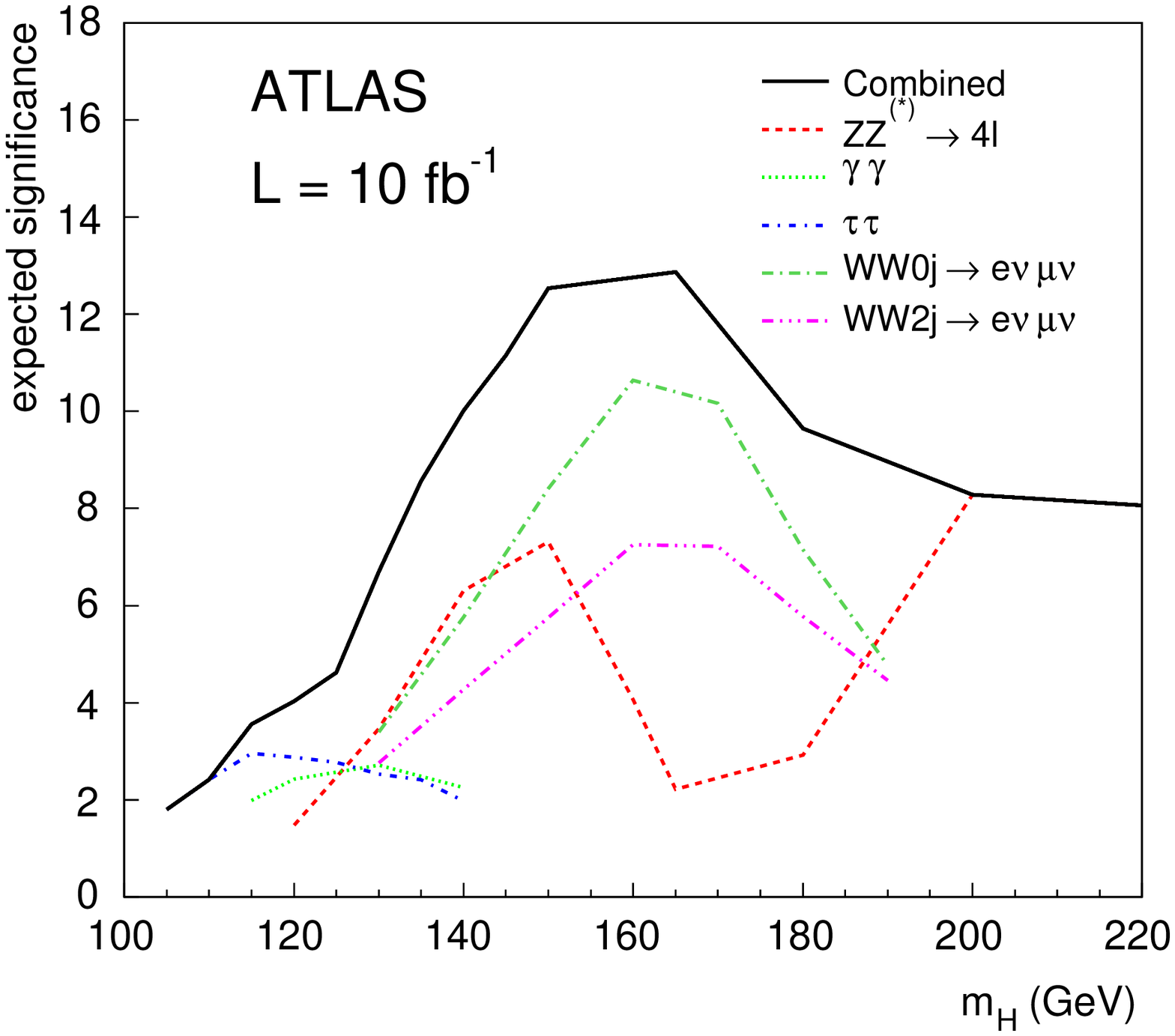,width=5.5cm, height=5.0cm}
  \epsfig{file=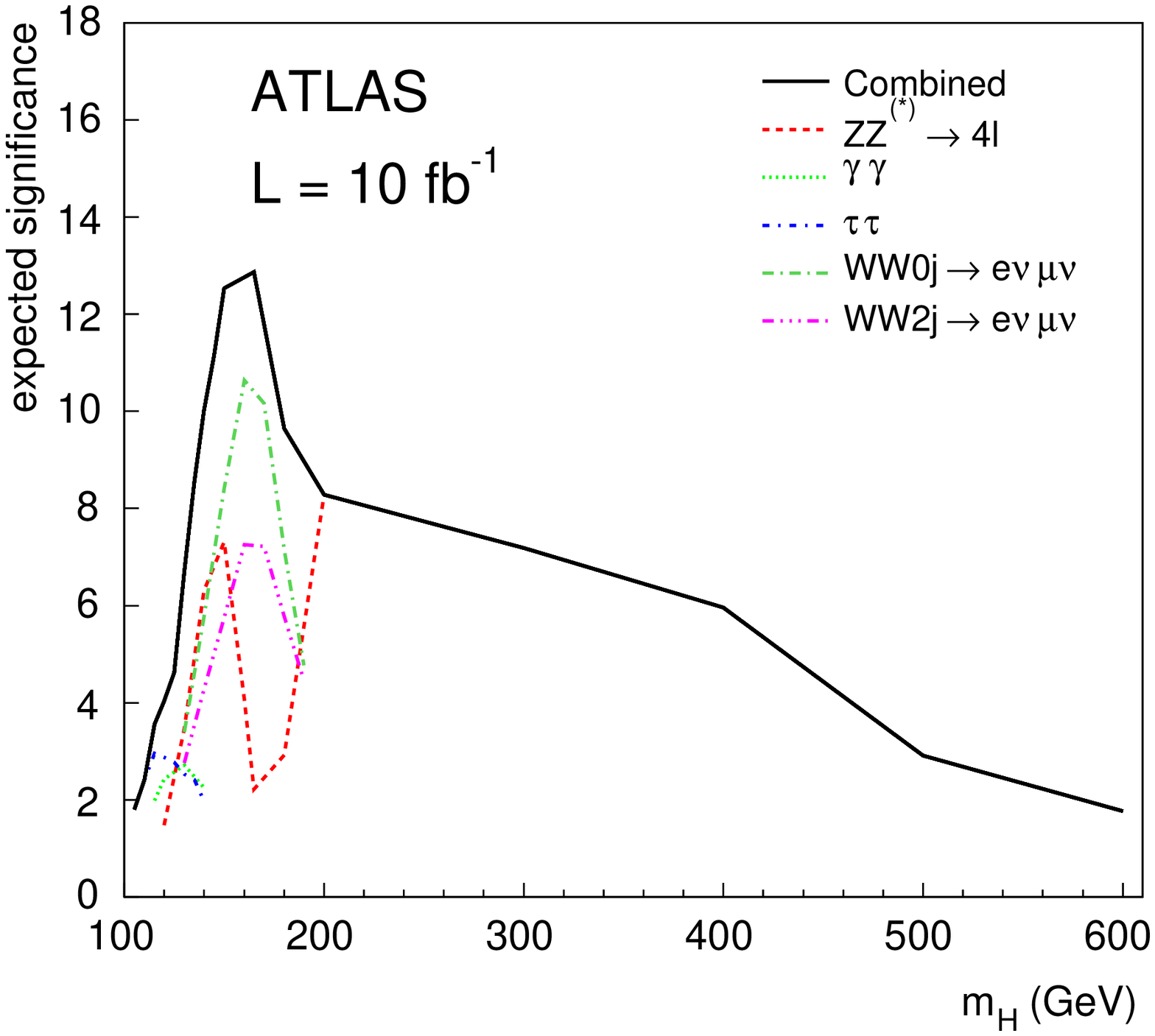,width=5.5cm, height=5.0cm}\\
}
\end{center}
\caption{\em The median discovery significance for the Standard Model Higgs Boson for the 
various channels as well as the combination for the integrated luminosity of 10 $fb^{-1}$
for (left) the lower mass range (right) for masses
up to 600 GeV. From Ref.~\cite{CERN-OPEN-2008-020}. 
\label{FS3.7}}
\end{Fighere}

\begin{Fighere}
\begin{center}
{
  \epsfig{file=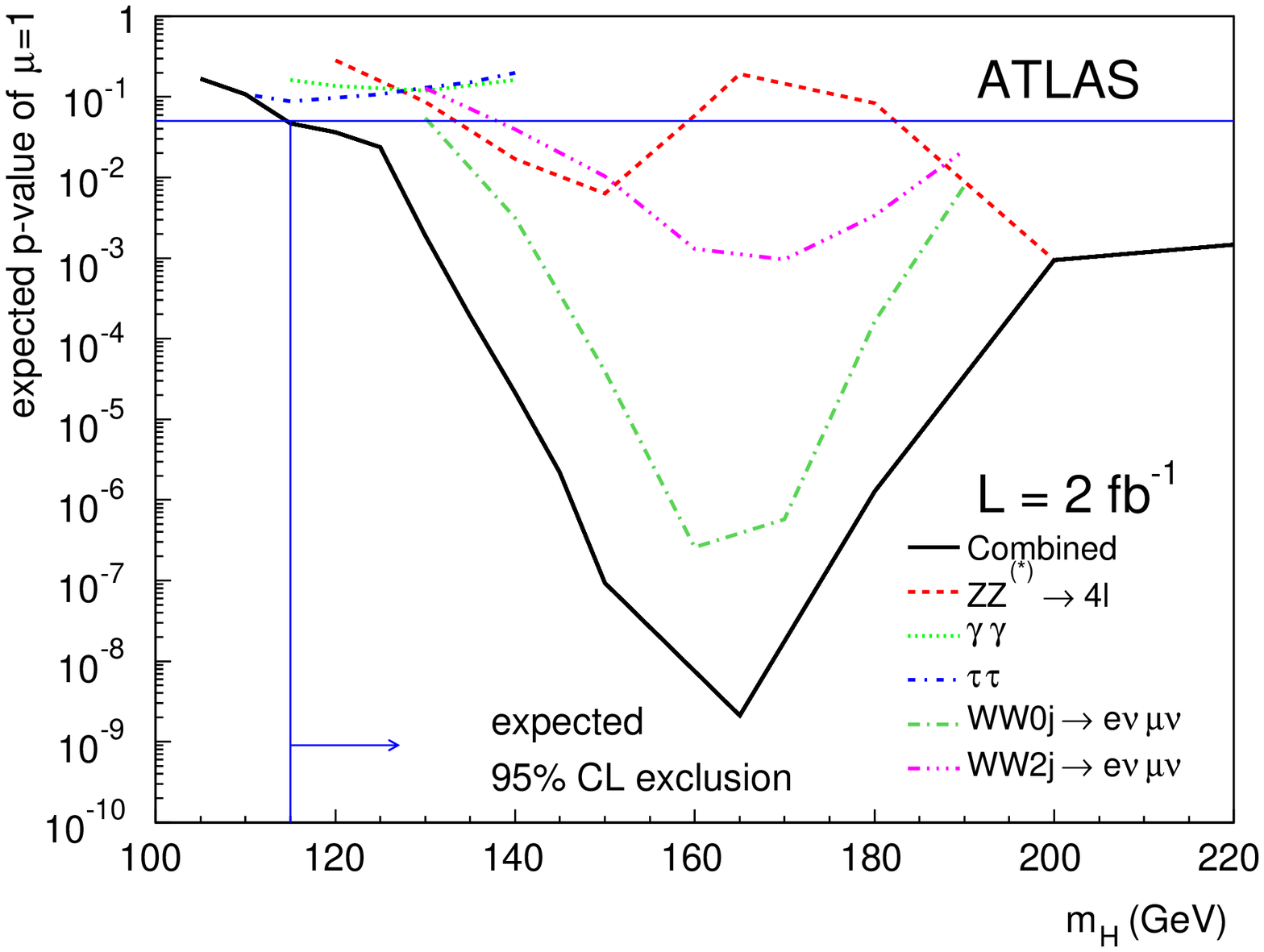,width=5.5cm, height=5.0cm}
  \epsfig{file=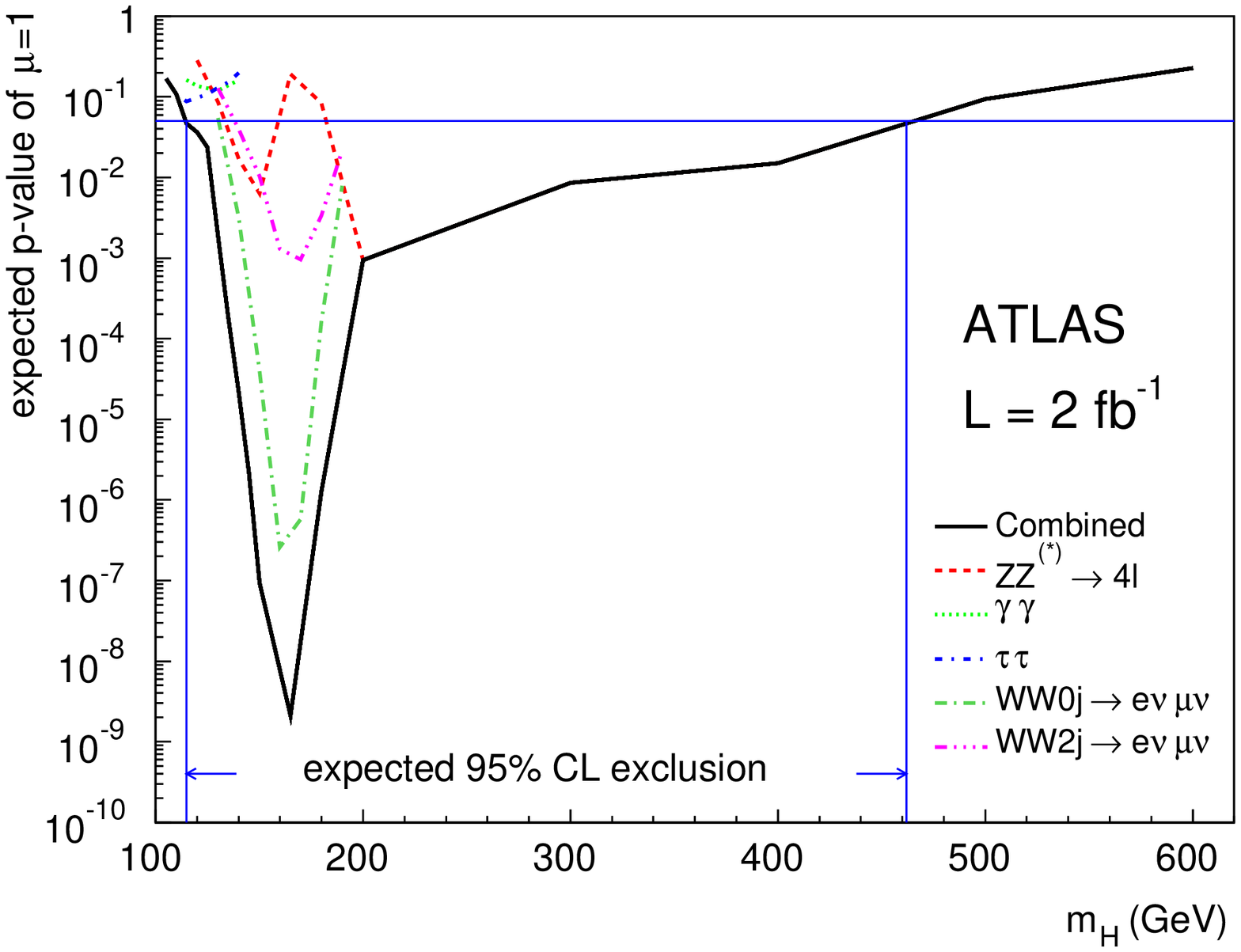,width=5.5cm, height=5.0cm}\\
}
\end{center}
\caption{\em The median  p-value obtained for excluding the Standard Model Higgs Boson for the 
various channels as well as the combination for (left) the lower mass range (right) for masses
up to 600 GeV. From Ref.~\cite{CERN-OPEN-2008-020}. 
\label{FS3.8}}
\end{Fighere}

\newpage
The procedure for the combination of search results based on the profile likelihood ratio has been 
applied to a study of the search for the Standard Model Higgs boson using four search channels:
$H \to \tau^+ \tau^-$, $H \to W^+ W^- \to e \nu \mu \nu$, $H \to \gamma \gamma$ and 
$H \to ZZ^{(*)} \to 4 \ell$. 
The resulting significances per channel and the combined one are shown in Fig.~\ref{FS3.7}
for  10~fb$^{-1}$ integrated luminosity. The median p-value obtained for excluding 
the Standard Model Higgs Boson for the various channels, as well as the combination for the 
lower mass range (left) and for masses up to 600~GeV (right) mass range, are shown 
in Fig.~\ref{FS3.8}. 

There is a decrease in expected significance for low $m_H$; however further developments of the
analysis will allow to incease the sensitivity in this region: for example 
improving analysis methods for $H \to \gamma \gamma$ channel with separation
of channels into zero or two accompanying jets topology.  Additional final states such as 
$ttH$ and $WH/ZH$ with $H \to b \bar b$ will help somewhat, 
although the contribution to the sensitivity will be small because of the large uncertainties 
in the background.  
For the $WW$ channel, the present studies include only $e \nu \mu \nu$ decay mode, but it is 
planned  to explore $e\nu e\nu$, $\mu \nu \mu \nu$ and $q q \ell \nu$ as well.
The additional $WW$ and $ZZ^{(*)}$ modes not discussed here, have been shown in past
to enhance sensitivity for a high mass Higgs.

The results obtained confirm the good discovery and exclusion sensitivities already shown 
in ATLAS Technical Design Report \cite{ATLASTDR}.
With a luminosity of 2 fb$^{-1}$ the expected (median) sensitivity is at the 5$\sigma$ level
or greater for discovery of the Higgs boson in the mass range 143-179~GeV, and the expected 
lower limit at 95\% confidence level on the Higgs mass is about 115~GeV.

\section{Prospects in the MSSM Higgs boson searches}

The LHC experiments have a large potential in the investigation of the MSSM Higgs sector. 
In the MSSM the couplings of Higgs bosons to fermions and bosons are different from those 
in the Standard Model resulting in different production cross-sections and decay rates.
While decays into ZZ and WW are dominant in the Standard Model, for Higgs masses above 
$m_{H} \sim 160$ GeV, in the MSSM these decay modes are either suppressed like 
$\cos(\beta - \alpha)$ in the case of the H (where $\alpha$ is the mixing angle of the two 
CP-even Higgs bosons) or even absent in case of the A. Instead, the coupling of the 
Higgs bosons to third generation fermions is strongly enhanced for large regions of the 
parameter space.

The search for light neutral Higgs bosons is based on the same channels considered for 
the Standard Model Higgs case. Heavier Higgs bosons will be searched for 
with the analysis of additional decay channels which become accessible in certain regions 
of the MSSM parameter space, due to enhanced couplings. Example are the  decay modes 
$H/A \to \tau \tau$ and $ H/A \to \mu \mu$, relevant for large values of $\tan \beta$. 
Decays into $\tau$ leptons also contribute 
to the search for charged Higgs bosons, which at the LHC can be extended to masses beyond 
the top-quark mass. 
The comprehensive study of the ATLAS detector discovery potential has been 
presented in \cite{ATLASTDR}. These studies have provided the first complete 
results for the ATLAS detector. Since then, they have been updated to include more studies 
on MSSM models~\cite{MSchumacherSUSY04}. 
In the most recent publication \cite{CERN-OPEN-2008-020}, more detailed analyses based 
on as-installed detector simulation as well as more recent Monte Carlo  generators 
have been revisited for the most important channels, with emphasis on the data-driven 
analysis of the background systematics and elaborate statistical methods for combination, 
as discussed in the Standard Model case. 

Below I recall, as present available, the preliminary picture for different benchmark 
scenarios, and in more details discuss analyses which have been revisited 
recently for $H^{\pm} \to \tau \nu$,  $H/A \to \tau \tau \to \ell \ell 4\nu$  
and $H/A \to \mu \mu$ searches.

\subsection{The MSSM discovery potential in various benchmark scenarios}

Different benchmark scenarios have been proposed for the interpretation of the MSSM Higgs 
boson searches \cite{MCarena}. In the MSSM, the masses and couplings of the Higgs bosons
depend, in addition to $\tan \beta$ and $m_A$, on the other parameters through 
radiative corrections. In particular, the phenomenology of the light Higgs boson h depends 
on the scenario and the following ones have been considered as representative ones:
\begin{itemize}
\item
{\it $m_h$-max}: the SUSY parameters are chosen in such way that for each point 
in the $(m_A, \tan \beta)$
parameter space the lightest Higgs boson mass $m_h$ close to the maximum possible value 
is obtained.
\item
{\it No-mixing scenario}: vanishing mixing in the stop sector is assumed, this scenario 
typically gives a small mass of the lightest Higgs boson h and is less favorable 
for the LHC.
\item
{\it Gluophobic scenario}: the effective coupling of the light Higgs to gluons is strongly
suppressed as the consequence of the strong mixing in the stop sector.
\item
{\it Small $\alpha$ scenario}: the parameters are chosen in such way that the effective mixing 
angle between the CP-even Higgs bosons is small. This results in the reduced branching ratio
into $b \bar b$ and $\tau \tau$ for large $\tan \beta$ and intermediate values of $m_A$. 
\end{itemize}

The overall discovery potential for the four scenarios is presented in Fig.~\ref{FS4.1}.
for the integrated luminosity of 30 fb$^{-1}$. The full parameter space can be covered 
for all benchmarks. However, in the region of moderate $\tan \beta$ and large $m_A$ only, 
one MSSM Higgs boson, the lightest Higgs boson (h) can be discovered. This remains true even for 
the 300 fb$^{-1}$  integrated luminosity. The exact location of that region depends
on details of the considered model.

\begin{Fighere}
\begin{center}
{
  \epsfig{file=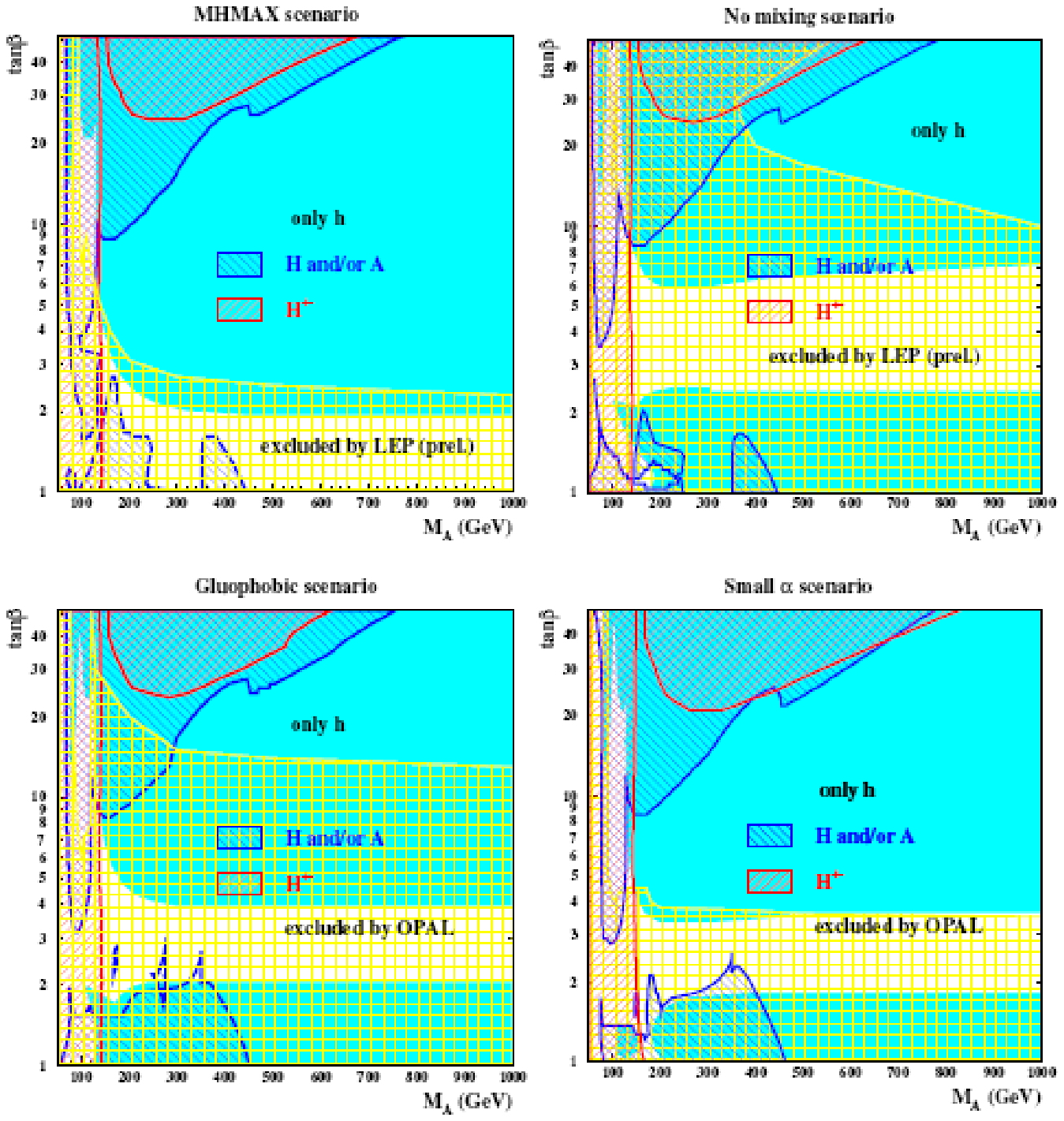,width=10.0cm, height=10.0cm}
}
\end{center}
\caption{\em Discovery potential for the Higgs bosons discovery in different benchmark 
scenarios (see text). In the blue area only lightest Higgs boson h can be observed. 
In the blue hatched area the heavy neutral Higgs bosons H and/or A, and in the red hatched 
area the charged Higgs bosons H$^{\pm}$ can be detected. The cross-hatched yellow region
is excluded by searches at LEP. ATLAS preliminary results \cite{MSchumacherSUSY04}.
\label{FS4.1}}
\end{Fighere}
\vspace{2mm}

For the above benchmark scenarios it was assumed  that CP is conserved in the Higgs sector.
Although CP conservation is present at Born level, CP violating effects might be introduced
via complex mixing parameters $A_t$, $A_b$ or mass terms. As a consequence, the mass eigenstates 
H$_1$, H$_2$  and H$_3$ are mixture of the CP eigenstates: h, H and A. 
A special scenario considered as 
a benchmark one is the so called CPX scenario \cite{Carena2000}, which is characterised by large 
phases. In this scenario,  uncovered by LEP region in parameter space exists for the H$_1$,
the lightest mass eigenstate \cite{MSchumacher06}. Preliminary studies have shown, 
that even at the LHC, uncovered regions at medium $\tan \beta \sim 5$ and small 
$m_{H^{\pm}}$ around 150 GeV which corresponds to small H$_1$  masses - remain, 
as illustrated in Fig.~\ref{FS4.1a}.
It needs to be studied further  if can be cover by exploiting additional search channels.
A similar situation may occur in the context of the next-to-minimal Standard Model (NMSSM).

\begin{Fighere}
\begin{center}
{
  \epsfig{file=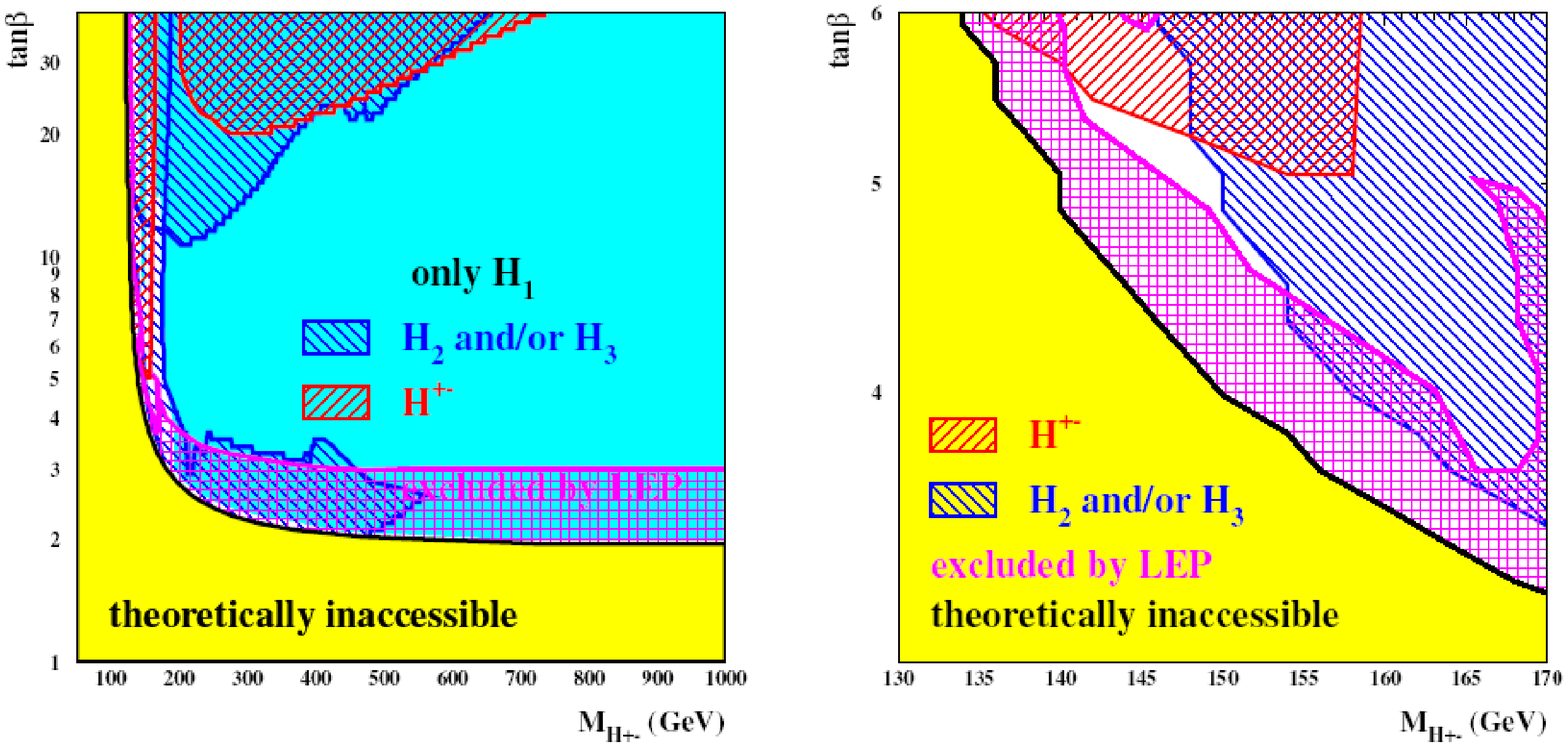,width=10.0cm, height=5.0cm}
}
\end{center}
\caption{\em 
Overall discovery potential for the Higgs bosons in the CPX scenario for data 
corresponding to the integrated luminosity of 300 fb$^{-1}$. 
The cross-hatched yellow region
is excluded by searches at LEP. ATLAS preliminary results \cite{MSchumacher06}.
\label{FS4.1a}}
\end{Fighere}

\subsection{Charged Higgs bosons}

The discovery of the charged Higgs boson would be a tangible proof of physics beyond 
the Standard Model. The recent studies present potential for discovery in search 
for five different final states arising from the three dominating fermionic decay modes.
The prepared search analyses cover the region below the top-quark mass, taking into
account present experimental constraints, the transition region with the charged Higgs 
boson mass of the order of top-quark mass, and the high mass region with the charged
Higgs boson mass up to 600 GeV. 

The sensitivities for discovery and exclusion are calculated with the profile likelihood method, 
combining all channels and including  systematic and statistical uncertainties.
The results are summarised in combined discovery and exclusion contours shown in Fig.~\ref{FS4.2}
for m$_h$-max scenario. The dependence of the H$^{\pm}$ discovery sensitivity on the specific choice 
of the MSSM parameter values is rather small, with the exception of the Higssino mixing parameter.
A significant improvement of present days constraints can already be achieved with limited 
but well understood data (less than 1 fb$^{-1}$).

\begin{Fighere}
\begin{center}
{
  \epsfig{file=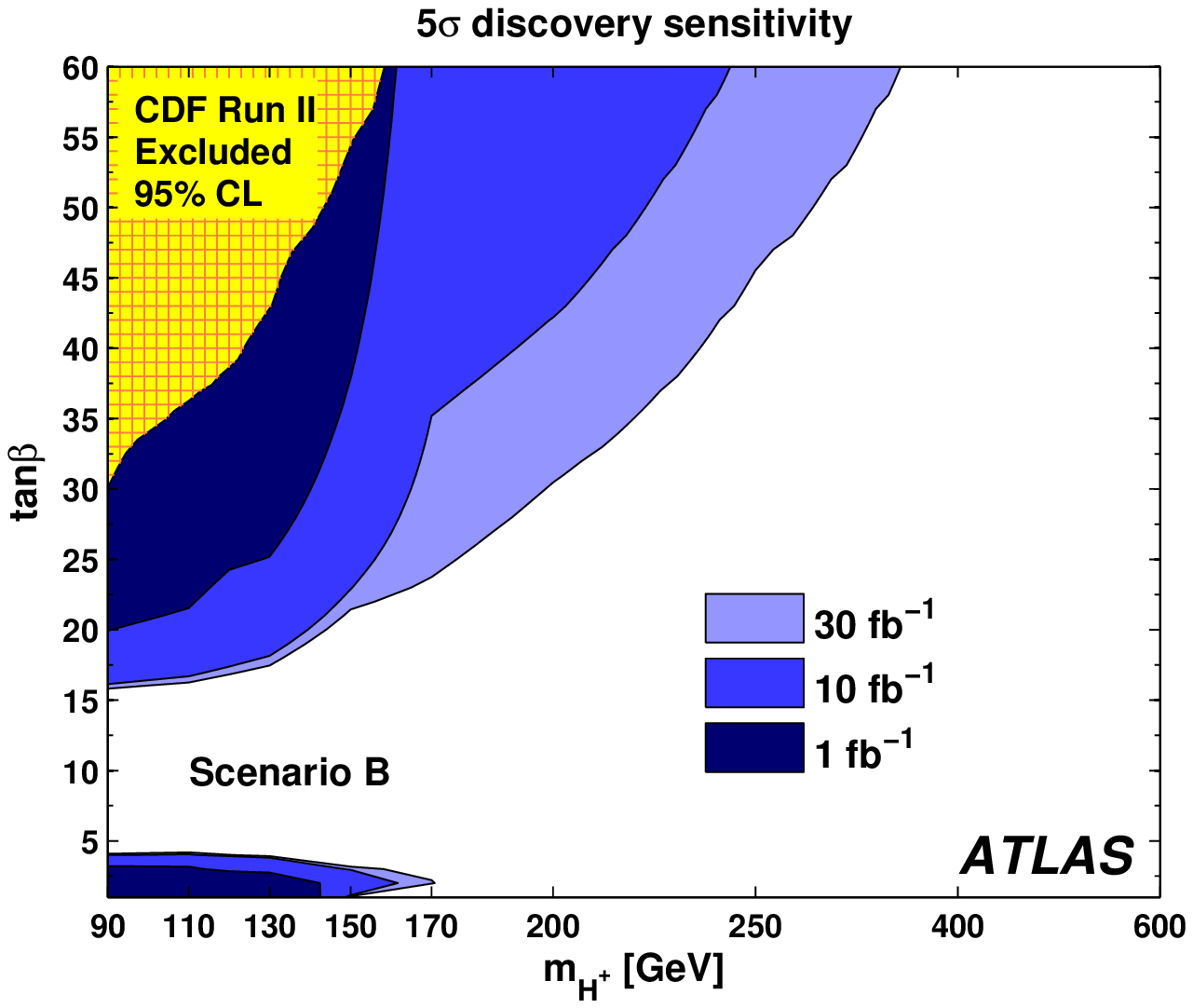,width=5.0cm, height=5.0cm}
  \epsfig{file=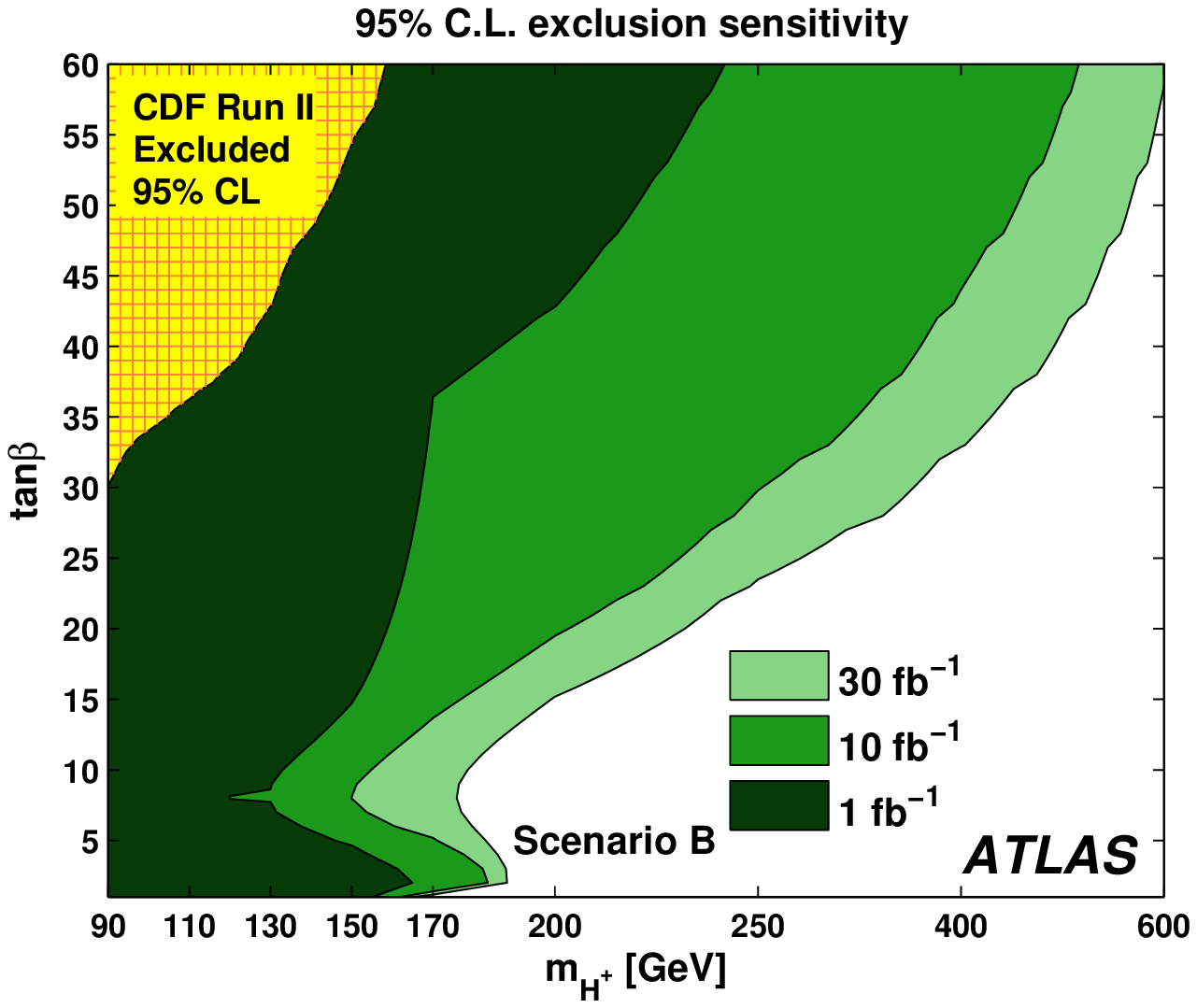,width=5.0cm, height=5.0cm}\\
}
\end{center}
\caption{\em Scenario B (mhmax): Combined Results. Left: Discovery contour, Right: 
Exclusion contour. Systematic and statistical uncertainties are included. 
From Ref.~\cite{CERN-OPEN-2008-020}. 
\label{FS4.2}}
\end{Fighere}
\vspace{2mm}

Below the top-quark mass the charged Higgs bosons are predominantly produced in top-quark decays
and the main decay mode is $H^+ \to \tau \nu$. Three different combinations of final states 
have been studied separately ($\tau$ and W bosons decaying leptonically and/or hadronically) each 
of them has shown to have potential for overperforming present sensitivity of Tevatron experiments 
already with 1 fb$^{-1}$. Analyses for adding also di-lepton final states are underway.
However, for the intermediate $\tan \beta$ $\sim$7, no discovery sensitivity is present,
although the production of this boson can be excluded in this region. 
This reduced sensitivity (vs previous estimates) 
represents rather conservative approach taken due to insufficient statistics available in MC samples
to estimate systematic uncertaintities, the sensitivity is recovered in the no-systematics limit.

The search for the heavy charged Higgs boson $(m_{H^{\pm}} > m_t)$ at Tevatron has been 
recently reported in \cite{D0arXiv:0807.0859v1}, see also review article \cite{arXiv:0903.0046}.
At the LHC, the main production mode will be the  $gb$ fusion $(gb \to tH^+)$. According to current 
expectations the discovery potential will be primarily in $\tau \nu$ decay channel 
even if the dominant mode is decay to $tb$ quarks. 
The discovery reached in $\tan \beta$ strongly depends on the Higgs boson mass.

In general, the ATLAS experiment will have sensitivity to explore the charged Higgs sector over sizable 
region of the parameter space with first 10~$fb^{-1}$. 
For a high SUSY mass scale, the charged Higgs boson could be the 
first signal of New Physics (and indication for Supersymmetry) discovered.

\subsection{The heavy Higgs bosons in $H/A \to \tau^+ \tau^-$ channel}

The coupling of the Higgs bosons to third generation fermions is strongly enhanced 
for large regions of the parameter space. The decay of the neutral Higgs bosons into pair of $\tau$ leptons
therefore constitutes an important discovery channel at LHC. The production of the Higgs bosons 
can proceed via two different processes; gluon-fusion or production in association with $b$ quarks.
The highest sensitivity is expected \cite{ATLASTDR} in the lepton-hadron final state of tau decays, 
and statistical combination of topologies with 0 b-jet (dominating medium $\tan \beta$) and one b-jet 
(dominating at higher $\tan \beta$ regions). 

\begin{Fighere}
\begin{center}
{
  \epsfig{file=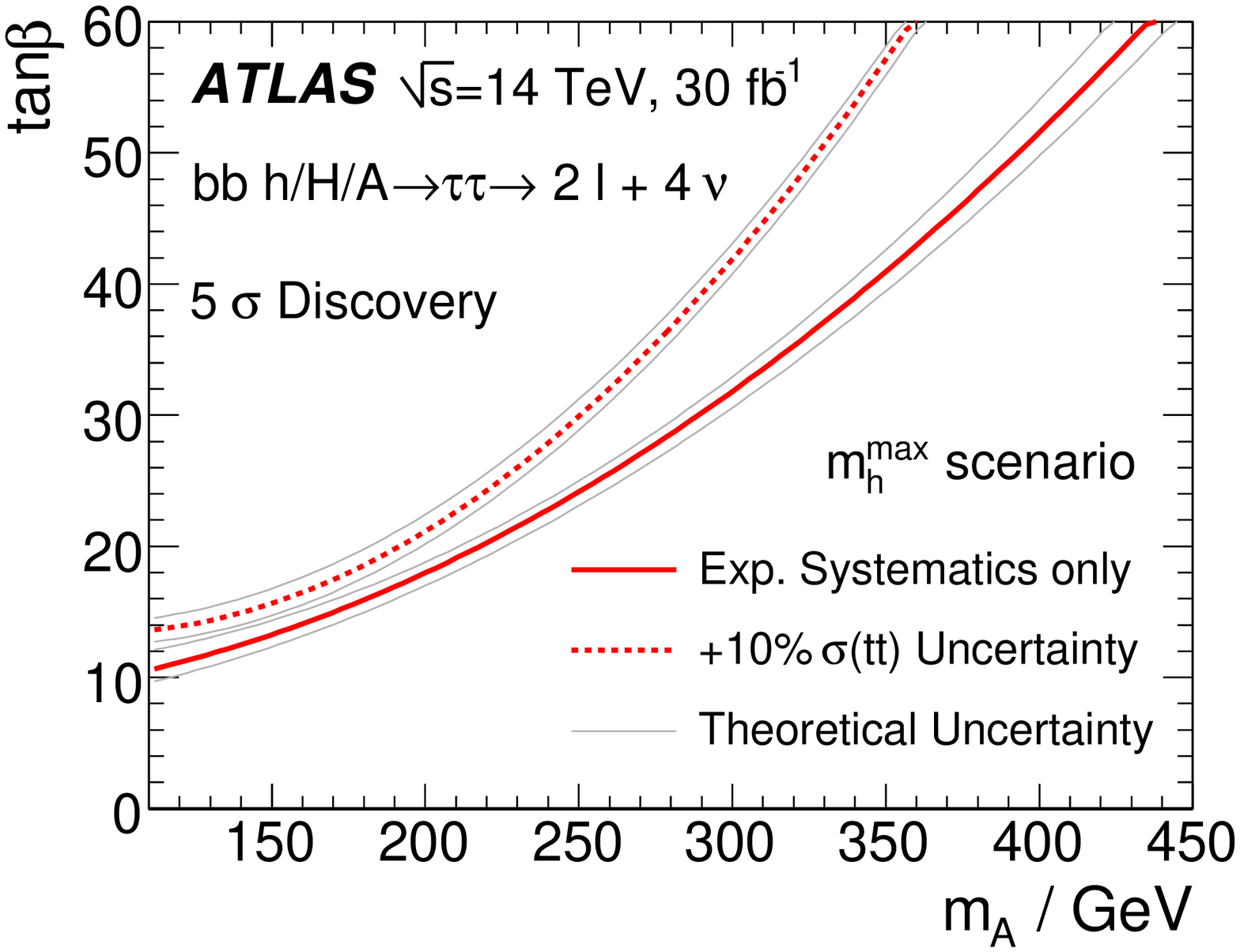,width=5.0cm, height=5.0cm}
  \epsfig{file=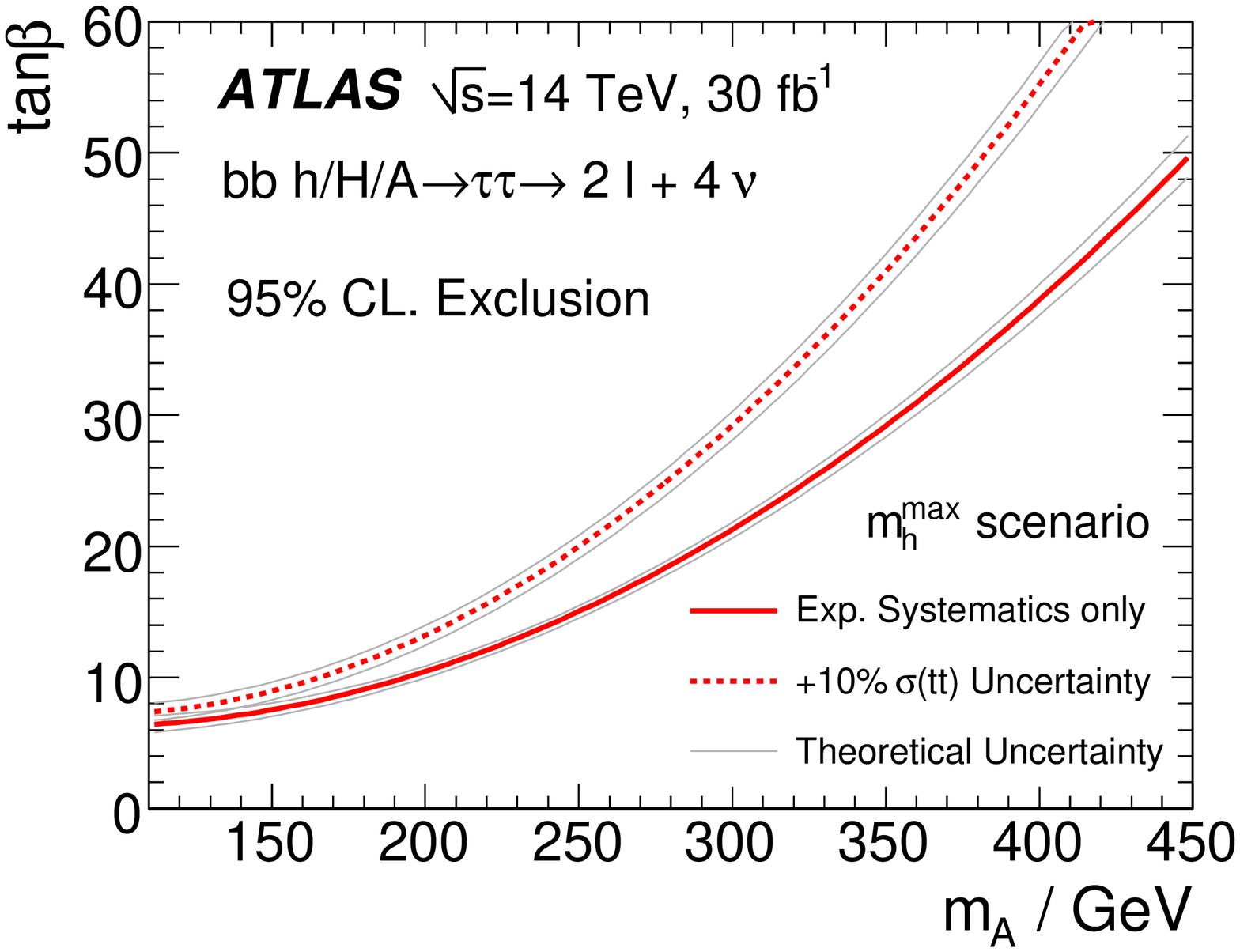,width=5.0cm, height=5.0cm}\\
}
\end{center}
\caption{\em The five $\sigma$ discovery potential (left) and the 95\% exclusion limit (right) 
as a function of $m_A$ and $\tan \beta$. The solid line represents the main result of the analysis. 
The dashed lines indicate the discovery potential and exclusion limit including an addition 
10\% uncertainty on the tt cross-section. The bands represent the influence of the systematic 
uncertainty on the signal cross-section. From Ref.~\cite{CERN-OPEN-2008-020}. 
\label{FS4.3a}}
\end{Fighere}
\vspace{2mm}

Recently, \cite{CERN-OPEN-2008-020}, detailed analysis for 
the $h/H/A \to \tau^+ \tau^- \to \ell^+ \ell^- 4 \nu$ final state, with at least 1 b-jet identified 
as coming from b-quark, has been performed. Particular emphasis was put on the data-driven procedure to 
estimate shape and normalisation of the resonant $Z \to \tau \tau$ background, dominant in low mass 
region, from measurements of the  $Z \to ee$ and $Z \to \mu \mu$ channels with the data. 
The expected discovery potential and exclusion limits are shown in Fig.~\ref{FS4.3a} for 30 fb$^{-1}$. 

Although the discovery potential of the lepton-lepton channel is weaker than in the lepton-hadron channel,
such measurement maybe easier at the beginning as does not require identification of the hadronically 
decaying $\tau$ leptons.

\subsection{The heavy Higgs bosons in $H/A \to \mu^+ \mu^-$ channel}

The decay of neutral MSSM Higgs bosons A, H and h into two muons is strongly enhanced
in the MSSM for large values of $\tan \beta$ (for the h if far from the decoupling limit),
and can be explored either as a discovery channel or for the exclusion of a large region of the 
$m_A - \tan \beta$ parameter space. Although the $\mu^+ \mu^-$ signature has substantially
smaller branching ratio if compared to $\tau \tau$ channel (scales as $(m_{\mu}/m_{\tau})^2$ )
it has the advantage of very clean signature in the detector, which allows also for a precise 
and direct Higgs boson mass measurement.
The event selection criteria are optimised in the signal mass range from 100 to 500 GeV, separately
for the signature with 0~b-jets (for the direct production in gluon fusion) and with at least
1~b-jet in the final state (for associated production with b-quarks, strongly enhanced 
at large $\tan \beta$).
\begin{Fighere}
\begin{center}
{
  \epsfig{file=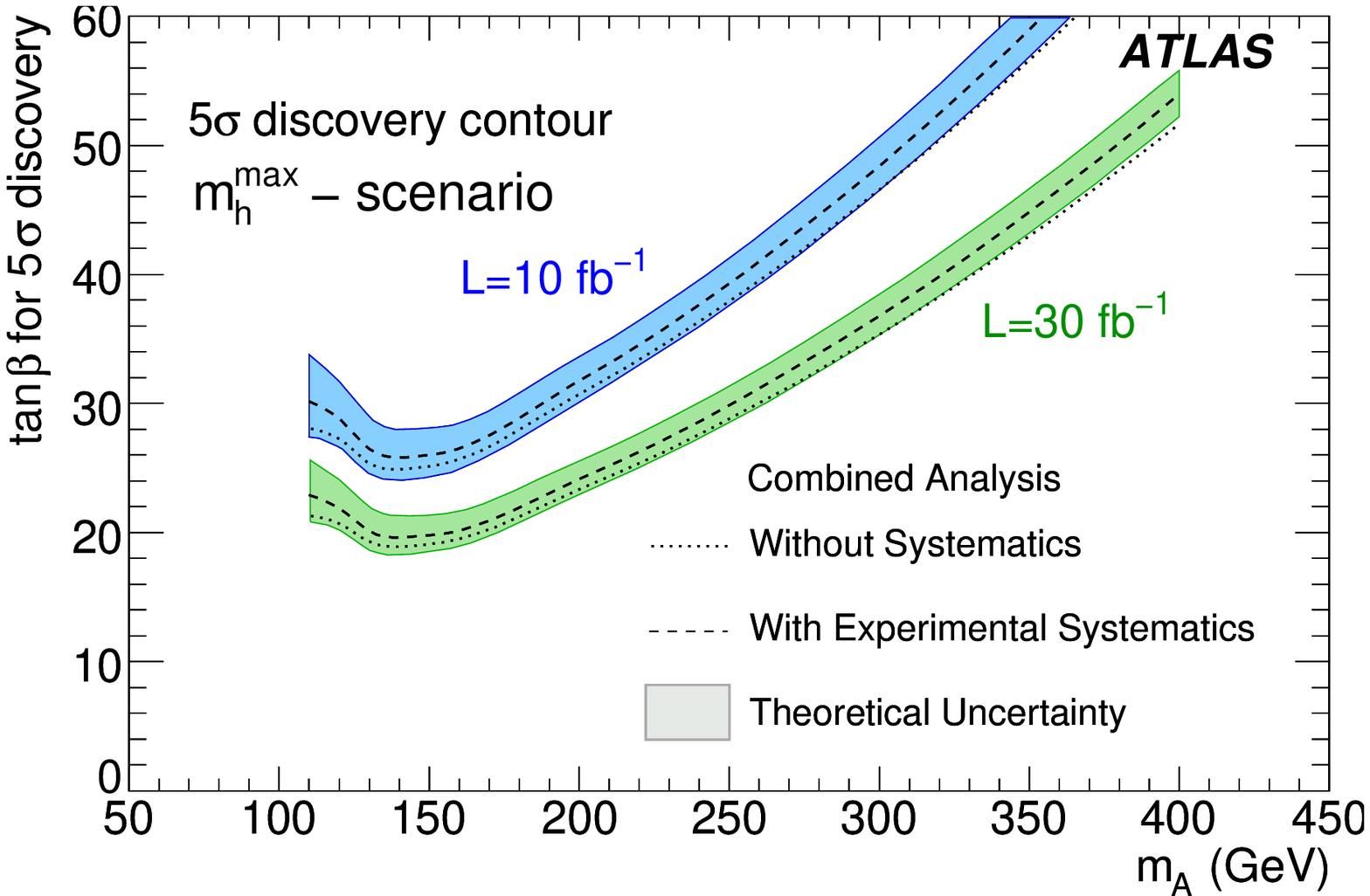,width=5.0cm, height=5.0cm}
  \epsfig{file=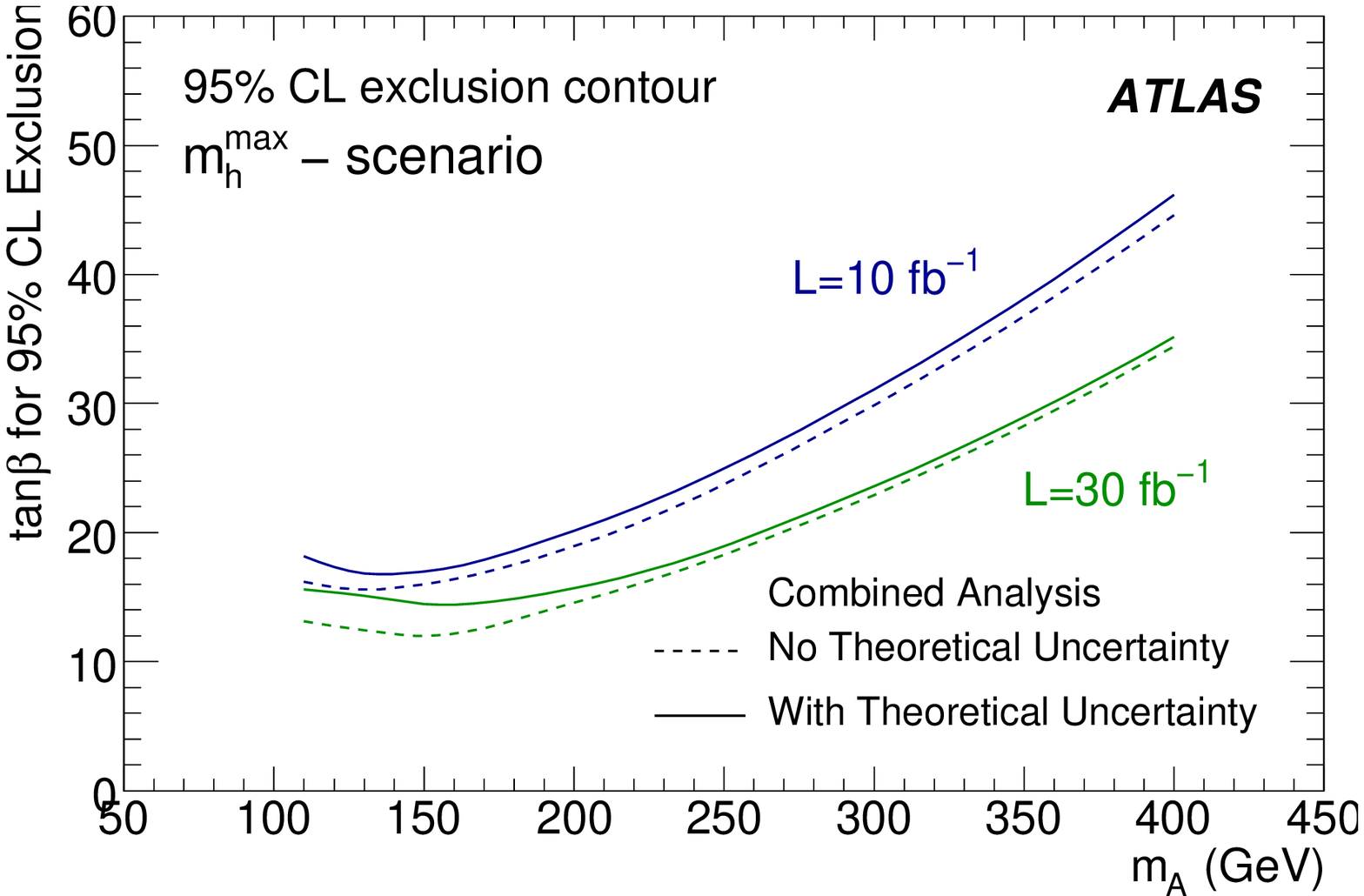,width=5.0cm, height=5.0cm}\\
}
\end{center}
\caption{\em Combined analyses results: (left) $\tan \beta$ values needed for the 5$\sigma$-discovery at 
10 fb$^{-1}$ and 30 fb$^{-1}$ of the integrated luminosity, shown is dependence on the A boson mass 
and (right) combined 95\% CL exlucion limits. From Ref.~\cite{CERN-OPEN-2008-020}. 
\label{FS4.3}}
\end{Fighere}
\vspace{2mm}

The obtained combined results, presented in Fig.~\ref{FS4.3} show that the integrated luminosity of 10 fb$^{-1}$ 
will allow for the discovery for m$_A$ masses up to 350 GeV with $\tan \beta$ values between 30-60. 
The three times higher luminosity allows to increase sensitivity down to $\tan \beta$ = 20.
The theoretical and detector-related systematic uncertainties are estimated to degrade the signal
significance by up to 20\%. Given the good mass resolution for di-muon final states, this channel 
can eventually be used to separate nearby Higgs boson resonances using a precise measurement of the 
shape of mass distributions.

\section{Summary}

The ATLAS experiment at LHC is well set up to explore the existence of the Standard Model or the MSSM Higgs 
bosons and are also prepared for unexpected scenarios. 

I have discussed several important channels for the Higgs boson searches, for which analyses have been 
recently revisited  as an important milestone of  preparing physics programme
for the ATLAS experiment with first 10-30 fb$^{-1}$ of the integrated luminosity.

With respect to previous estimates, analyses have been completed with very detailed studies on the 
theoretical and detector related systematics, also the data-driven methods for controlling background 
shapes and normalisation have been established in most cases. The most recent Monte Carlo generators 
and  NLO cross sections have been used for signal and backgrounds if available. In addition, 
the very advanced statistical procedures to estimate discovery and exclusion reach have been prepared 
and their performance evaluated on the basis of Monte Carlo studies.

In general, recent studies reported  confirm good discovery  
sensitivities of ATLAS detector for the Standard Model and the MSSM Higgs boson searches.
The full Standard Model mass range and the full MSSM parameter
space can be covered (for CP conserving case).

The LHC data will hopefully give soon guidance to the theory and to the future high energy physics 
experiments.

{}


\begin{thebibliography}{}

\bibitem{EWModel}
S. L. Glashow, Nucl. Phys. {\bf 22} (1961) 579; S. Weinberg, Phys. Rev. Lett. {\bf 19} (1967) 1264;
A. Salam, in Elementary Particle Theory, ed. N. Svartholm, Stockholm, {\em ``Almquist and Wiksell''}
(1968), 367.

\bibitem{QCDModel}
H. D. Politzer, Phys. Rev. Lett. {\bf 30} (1973) 1346; 
D. J. Gross and F. E. Wilczek, Phys. Rev. Lett. {\bf 30} (1973) 1343; H. Fritzsch, 
M. Gell-Mann and H. Leutwyler, Phys. Lett. {\bf B47} (1973) 365.

\bibitem{SMHiggs}
P. W. Higgs, Phys. Rev. Lett. {\bf 12} (1964) 132 and Phys. Rev. {\bf 145} (1966) 1156;
F. Englert and R. Brout, Phys. Rev. Lett. {\bf 13} (1964) 321;
G. S. Guralnik, C. R. Hagen and T. W. Kibble, Phys. Rev. Lett. {\bf 13} (1964) 585.

\bibitem{SM&MSSM}
For a review see for example: J. F. Gunion, H. E. Haber, G. Kane and S. Dawson, {\it The Higgs Hunter's 
Guide}, Frontiers in Physics Series (Vol. 80), Addison-Wesley publ., ISBN 0-201-50935-0; 
H. P. Nilles, Phys. Rep. {\bf 110} (1984) 1; H. E. Haber and G. L. Kane, Phys Rep. {\bf 117} (1985) 75;  

\bibitem{UnitStabil}
L. Maiani, G. Parisi and R. Petronzio, Nucl. Phys. {\bf B136} (1979) 115; N. Cabbibo et al., 
Nucl. Phys. {\bf B158} (1979) 295;  G. Altarelli and G. Isidori, Phys. Rev. Lett. {\bf B337} (1994) 141;
J. A. Casas, J. R. Espinosa and M. Quiros, Phys.lett. {\bf B342} (1995) 171, Phys. Lett. {\bf B383} (1996) 374; 

\bibitem{LEPlimit}
ALEPH, DELPHI, L3 and OPAL Collaborations, Phys. Lett. {\bf B565} (2003) 61. 

\bibitem{EWFit}
LEP Electroweak Working Group, July 2008, http://lepewg.web.cern.ch/LEPEWWG.

\bibitem{Tevatron}
The TEVNPH Working Group (for the CDF and D0 Collaboration), 
{\it Combined CDF and Dzero Upper Limits on Standard Model Higgs Boson Production 
 with 4.2 fb$^{-1}$ of Data}, FERMILAB-PUB-09-060-E, CDF Note 9713, D0 Note 5889.
 
\bibitem{MSSMLEP}
By DELPHI Collaboration (J. Abdallah et al.), Eur.Phys.J.{\bf C54} (2008) 1, Erratum-ibid. {\bf C56} (2008) 165.

\bibitem{MSSMChrgatTeV}
A. Abulencia et al., The CDF Collaboration, Phys. Rev. Lett. 96, 042003 (2006),
V. Abazov et al., D0 Collaboration, arXiv.org:0807.0859.   

\bibitem{MSSMatTeV}
A. Abulencia et al., The CDF Collaboration, Phys. Rev. Lett. {\bf 96}, 011802 (2006).
 V. Abazov et al., D0 Collaboration, Phys. Rev. Lett. {\bf 102}, 051804 (2009)

\bibitem{CERN-OPEN-2008-020}
The ATLAS Collaboration, {\it Expected Performance of the ATLAS Experiment,
Detector, Trigger and Physics}, CERN-OPEN-2008-020, Geneva, 2008, arXiv:0901.0512.

\bibitem{ATLASTDR}
The ATLAS Collaboration, {\it Detector and physics performance technical design report}, 
CERN/LHCC/99-15, 1999.


\bibitem{ATLASDetPaper}
The ATLAS Collaboration (G. Aad et al.), {\it  The ATLAS Experiment at the CERN Large Hadron Collider},
Published in JINST 3:S08003,2008.

\bibitem{Jakobs-EurJPhysC} 
K. Jakobs, {\it Higgs bosons at the LHC}, Eur. Phys. J. C DOI 10.1140/epjc/s10052-008-0746-8.


\bibitem{MCarena}
M. Carena, S. Heinemeyer, C. E. Wagner, G. Weiglein, Eur. Phys. J. {\bf C26}, 601 (2003).

\bibitem{MSchumacherSUSY04}
M. Schumacher, {\it Investigation of the Discovery Potential for Higgs Bosons of the Minimal
Supersymmetric Extension of the Standard Model with ATLAS}, hep-ph/0410112,
presented at SUSY04 conference, October 8th 2004.

\bibitem{Carena2000}
M. Carena et al., Phys. Lett. {\bf B495}, 155 (2000).

\bibitem{MSchumacher06}
E. Accomando et al. {\it Workshop on CP Studies and Non-Standard Higgs Physics}, 
Report of the Workshop on CP Studies and Non-standard Higgs Physics, CERN, Geneva, 
Switzerland, May 2004 - Dec 2005, e-Print: hep-ph/0608079 

\bibitem{D0arXiv:0807.0859v1}
D0 Collaboration (V. Abazov et al.), {\it  Search for charged Higgs bosons decaying 
to top and bottom quarks in ppbar collisions}, arXiv:0807.0859v1 [hep-ex].

\bibitem{arXiv:0903.0046}
J. L. Feng, JF Grivaz, J. Nachtman, {\it Searches for Supersymmetry at High-Energy Colliders},
arXiv:0903.0046v1 [hep-ex].


\end{thebibliography}
\end{document}